\documentclass[english]{article}
\usepackage[T1]{fontenc}
\usepackage[utf8]{inputenc}
\usepackage{geometry}
\usepackage{amsmath}
\usepackage{amssymb}
\usepackage{graphicx}
\usepackage[authoryear, round]{natbib} 
\usepackage{float}
\usepackage[caption=false]{subfig}

\usepackage{setspace}

\makeatletter

\providecommand{\tabularnewline}{\\}

\@ifundefined{date}{}{\date{}}

\usepackage[usenames,dvipsnames]{color}
\usepackage{amsthm}\usepackage[english]{babel}
\usepackage{tabularx}
\usepackage{multirow}
\usepackage{enumerate}
\newcommand{\rbr}[1]{\left( #1 \right)}
\newcommand{\sbr}[1]{\left[ #1 \right]}
\newcommand{\cbr}[1]{\left\{ #1 \right\}}
\newcommand{\R}{\mathbb{R}}

\theoremstyle{plain}
\newtheorem{definition}{Definition}[section]
\newtheorem{theorem}{Theorem}[section]

\newtheorem{lemma}{Lemma}
\newtheorem{corollary}{Corollary}[section]

\makeatother
\usepackage{enumitem}

\usepackage{babel}
\begin{document}
\global\long\def\intr{\int_{R}}
 \global\long\def\cbr#1{\left\{  #1\right\}  }
 \global\long\def\rbr#1{\left(#1\right)}
 \global\long\def\ev#1{\mathbb{E}{#1}}
 \global\long\def\R{\mathbb{R}}
 \global\long\def\norm#1#2#3{\Vert#1\Vert_{#2}^{#3}}
 \global\long\def\pr#1{\mathbb{P}\rbr{#1}}
 \global\long\def\cleq{\lesssim}
 \global\long\def\ceq{\eqsim}
 \global\long\def\conv{\rightarrow}
 \global\long\def\TDD#1{\text{{\color{red}ToDo:#1}}}
 \global\long\def\Red#1{\text{{\color{red}#1}}}
 \global\long\def\PM#1{\text{{\color{red}PM:#1}}}
 \global\long\def\Var#1{\text{Var}(#1)}
 \global\long\def\dd#1{\textnormal{d}#1}
 \global\long\def\inti{\int_{0}^{\infty}}
 \global\long\def\ODP#1{\text{{\color{green}#1}}}

\title{Ranking Policies Under Loss Aversion and Inequality Aversion}
\author{Martyna Kobus\footnote{Institute of Economics, Polish Academy of Sciences, mkobus@inepan.waw.pl. Martyna Kobus acknowledges receipt of the grant from the National Science Centre, Poland, No. 2020/39/B/HS4/03131.}, Radosław Kurek\footnote{Institute of Economics, Polish Academy of Sciences, radek.kurek@inepan.waw.pl.}, Thomas Parker\footnote{Department of Economics, University of Waterloo, tmparker@uwaterloo.ca.}}
\date{}
\maketitle
\begin{abstract}
Strong empirical evidence from laboratory experiments, and more recently from population surveys, shows that individuals, when evaluating their situations, pay attention to whether they experience gains or losses, with losses weighing more heavily than gains. The electorate’s loss aversion, in turn, influences politicians’ choices. We propose a new framework for welfare analysis of policy outcomes that, in addition to the traditional focus on post-policy incomes, also accounts for individuals’ gains and losses resulting from policies.
We develop several bivariate stochastic dominance criteria for ranking policy outcomes that are sensitive to features of the joint distribution of individuals’ income changes and absolute incomes. The main social objective assumes that individuals are loss averse with respect to income gains and losses, inequality averse with respect to absolute incomes, and hold varying preferences regarding the association between incomes and income changes. We translate these and other preferences into functional inequalities that can be tested using sample data. The concepts and methods are illustrated using data from an income support experiment conducted in Connecticut.
\end{abstract}
	
~
~\\	
\textbf{JEL class:} D04, C14

\noindent  \setcounter{page}{1}  

\section{Introduction}\label{sec:intro}

When assessing feasible policy changes, policy makers frequently face the challenge of balancing benefits for those who gain against the setbacks experienced by those who lose. A key insight from the behavioral economics literature \citep[e.g.][]{TverskyKahneman91} is that individual welfare is dependent not only on the current state but also on the change in states. Prospect theory, originally developed to explain individual choices in uncertain situations but soon extended to riskless settings \citep{Thaler80} posits that people base their decisions on whether they perceive the choice as leading to a gain or a loss\footnote{This depends on the reference point and of course stands in contrast to the assumptions of consistency or invariance in preferences made in rational choice theory.}, and that losses carry more weight than equivalent gains (i.e. loss aversion) \citep{KahnemanTversky79}. 

Loss aversion is today one of the best-documented concepts in the literature. A global study published in \textit{Nature Human Behaviour} \citep{Ruggierietal20} repeats Kahneman and Tversky's original 1979 experiment in 19 countries and 13 languages and confirms that it is broadly replicable. More recently, new evidence has emerged based on large representative population samples rather than laboratory experiments. \citet{BlakeCannonWright21} confirm that respondents in a UK survey are loss averse. They also confirm another prediction of the theory, namely that individuals are inequality averse for gains but inequality loving (or risk-seeking) for losses.\footnote{This is called the \textit{S-shapedness} of the value function and we use this property in Section \ref{sec:secondorder}.} These observations hold not only for the full sample, but also for every subsample of the data: for both men and women, at any age, income or level of education. Similarly, based on a large demographically representative survey in eight European countries, \citet{MeissnerGassmanFaureSchleich23} find that respondents are, on average, loss averse and weigh losses about twice as much as gains of the same size, an estimate that is consistent with the result of a meta-analysis of 150 laboratory and field studies by \citet{BrownImaiVieiderCamerer24}. Higher values are found for France, Italy, Sweden, Germany and the UK, and lower values for Poland, Romania and Spain. \citet{ChapmanSnowbergWangCamerer24} confirm the asymmetry between gains and losses in the preferences of the US population.\footnote{On the other hand, they also find a higher prevalence of loss tolerance in the population than in laboratory experiments.}


Given such strong empirical evidence, loss aversion has been invoked to explain a wide range of phenomena such as the endowment effect \citep{Thaler80}, the status quo bias \citep{SamuelsenZeckhauser88}, labor supply \citep{Camereretal97, Dunn96, KosegiRabin06} and job search \citep{DellaVignaetal17}, the equity premium puzzle \citep{BenartziThaler95}, tax evasion \citep{DhamialNowaihi10, Engstrometal15, ReesJones18}, price setting by firms \citep{Ahrensetal17}, incumbency biases in elections \citep{QuattroneTversky88}, consumption decisions \citep{Bowmanetal99, KosegiRabin06}, the timing of retirement \citep{Seibold21}, and many others. It is thus natural that loss aversion, as an inherent feature of human preferences, should be incorporated into the welfare analysis of policy outcomes. 

Moreover, loss aversion ought to be considered important for political economy reasons as well. Mounting evidence indicates that loss aversion of the electorate determines policy choices. By incorporating it into a model of trade policy determination, \citet{FreundOzden8} are able to explain why industries experiencing losses are more likely to be protected and why existing protectionist policies are persistent. In their framework, as in this paper, society is made up of individuals who exhibit loss aversion and the government takes this into account when maximising social welfare. Consequently, the government is concerned about constituencies who would suffer losses from a policy change. In a similar setting, \citet{Tovar9} finds that loss aversion of the electorate, if large enough, may be the reason for the anti-trade bias puzzle \citep{Rodrik95}.  \citet{AlesinaPassarelli19} show that voters' preference reversals towards policies such as the Affordable Act, the Smoke Free Air Act or carbon taxes can be explained by loss aversion. Rules and institutions that are overly protective and difficult to change may be designed by legislators due to the population's loss aversion \citep{AttanasiCorazziniPassarelli17}.

Beyond the need to account for an important aspect of individual preferences, and aside from political economy considerations, policy makers may also have purely normative reasons for incorporating loss aversion into policy evaluation. As an example of this, the recent pandemic sparked a significant debate on the ethical principles guiding vaccine development \citep{Kahnetal20, Solbakketal21}. \citet{Eyal20} argues that the Hippocratic maxim of ‘first, do no harm’ prompted US and EU policy makers to postpone the development of Covid-19 vaccines by refusing to authorize human challenge trials, in which a small group of volunteers would have been deliberately exposed to the virus. In this choice, policy makers prioritized the risk of harm to a few individuals over the potentially enormous gains from speeding up the delivery of vaccines to broad segments of society. 

\citet{Firpoetal} develop methods to rank policies in a way that is sensitive to individuals’ loss aversion. However, their framework assumes that individuals do not care about incomes at all and focus exclusively on income changes. It seems more realistic to assume that both dimensions matter. Indeed, the behavioral economics literature postulates \citep[see, e.g.][]{KosegiRabin06} that the extended utility function—commonly referred to as the value function—depends both on outcomes and on gains and losses.\footnote{A common functional form is the sum of a function that depends on outcomes and a function that depends on gains/losses \citep[see][for a review]{ODonoghueSprenger18}. This form is a special case of the classes of value functions considered in this paper (see Corollary \ref{corr:additive}).} Moreover, a policy that is optimal with respect to gains and losses may redistribute income in a way that perpetuates existing income inequality to socially unacceptable levels. Thus, accounting for inequality aversion with respect to incomes, the traditional concern of welfare analysis, remains indispensable. 

We thus significantly extend the framework considered in \citet{Firpoetal}. Namely, we consider a bivariate setting in which not only income changes matter, but also final incomes. A policy generates a distribution of incomes and, since it is always preceded by another distribution, it also generates a distribution of gains and losses. The standard approach in the welfare analysis \citep{Atkinson70} considers incomes alone. The approach of \citet{Firpoetal} considers changes only. We combine the two to obtain a more realistic setting for individual and policymaker preferences. This can alter the ranking of distributions obtained from considering only a single dimension. For example, suppose one distribution appears preferable in terms of losses, but primarily because the losses of richer individuals are smaller. Once income inequality aversion is taken into account, that distribution may no longer be preferred; the alternative distribution may be even favored over some range of values (of incomes and of income gains/losses). Conversely, consider a tax reform that reduces inequality—an outcome desirable for any inequality-averse decision maker. What still matters is how this reduction is achieved. If it results from substantial gains accruing to a few winners at the lower end of the income distribution, while many others in the same income range experience small losses, then the inequality reduction effect appears less persuasive. The proportions of winners and losers, and the magnitudes of their respective gains and losses, matter also for the political economy reasons already discussed. The bivariate joint framework makes it possible to capture not only the policy maker’s preferences for equality and loss aversion per se, but also preferences regarding how the two interact. For example, the policy maker may prefer policies for which losses are concentrated among high-income individuals rather than among those with low incomes. This highlights dependence, a distinctive feature of multivariate frameworks as opposed to univariate ones. Overall, adopting a bivariate perspective substantially broadens the scope of policy welfare comparisons.

We follow the social choice tradition of ranking the welfare resulting from policies by means of a social welfare function. In line with this literature \citep[see e.g.][]{Dalton20, Sen70, GajdosWeymark12}, the welfare function expresses the preference of a social decision maker who uses a utility function to transform individual outcomes into an interpersonally comparable measure of well-being. To take account of certain regularities in individual preferences, the utility function reflects qualitative properties such as the diminishing marginal utility of income or, in the context of this paper, loss aversion. We are therefore concerned with classes of utility functions and of social welfare functions. To be more precise, we refer to the class of welfare functions as a social value function, because instead of an individual utility function depending on income, there is an extended utility function (as mentioned, called a `value function') depending on both absolute income and income gains/losses. Such extended utility functions, with specific functional forms, are used in models of reference-dependent preferences \citep[e.g.][]{KosegiRabin06, ODonoghueSprenger18}. 

In such a setting, we develop bivariate dominance criteria to rank joint distributions of income and income gains and losses. These criteria are called dominance criteria because they hold for whole classes of value functions, thus making the ranking of policies or distributions robust. The theorems below link value functions from a specific class to a functional inequality that can be tested with observable features of the distribution of income and income changes. Such results take us from the realm of the unobservable (utility or value functions) to the realm of implementable conditions.

In standard risk and inequality comparisons, if the utility function is non-decreasing and concave, expected utility or utilitarian welfare is higher in an income distribution that stochastically dominates another distribution at the second order \citep{RothschildStiglitz70, Atkinson70}. Similarly, in the main result of the paper (Theorem \ref{thm:liasd}), we assume another set of conditions for the value function: that it is loss averse, inequality averse for incomes, and has other properties that can be linked to the so called higher-order risk preferences known from the literature on risk measurement \citep{EeckhoudtSchlesinger13}, The literatures on the measurement of risk and inequality measurement share a common analytical structure and concepts and results in one literature are used in the other \citep{RothschildStiglitz70, Atkinson70, GajdosWeymark12}. The first property is an aversion to positive associations between income and income changes, which relates to correlation aversion in risk measurement \citep{EeckhoudtReySchlesinger7}. It expresses a preference for less positively correlated incomes and income changes. The bivariate distribution in which there is higher likelihood for low (high) outcomes to be paired with large gains (high losses) is preferred to the distribution when the opposite happens. The next property is a preference for the association aversion to be larger at the bottom of the distribution (for losses and low incomes) than at the top (for high gains and high incomes). This is equivalent to cross-prudence in risk \citep{EeckhoudtReySchlesinger7}. Here it means that a policy maker's preference for equality in one dimension is stronger the lower the value of the other dimension. He/She prefers more equal outcomes among those who have lost than among those who have gained and a more equal distribution of gains among the poorer than among the richer. Final property is related to cross-temperance in risk preferences \citep{EeckhoudtReySchlesinger7}. Here it means that a policy maker's preference for equality in one dimension is stronger the lower the degree of equality of the other dimension.  

Given this set of qualitative features, expected value is higher in the distribution that dominates in terms of a criterion --- a specified set of testable conditions --- based on the joint distribution of income and income change induced by a policy. These qualitative features of value functions arise from the joint nature of our setting. This is also reflected in the dominance criteria, which combine Firpo et al.'s criterion for the univariate distribution of gains/losses, second-order stochastic dominance for the univariate distribution of income, and a dominance criterion for the integral of the joint distribution. The opposite result, on the other hand, uses the class of value functions that reverse the sign of the mentioned higher-order preferences and is given in Theorem \ref{thm:liasd2}.

Apart from the main result, which involves both loss aversion and inequality aversion, we also develop other results. In Theorem \ref{thm:lasbd} we combine loss aversion with first-order stochastic dominance so that a better distribution is the one that has lower losses/higher gains and at the same time higher incomes (including higher mean incomes). It also has a lower association between gains/losses and incomes. The opposite result, favoring a higher association, provides a related dominance criterion based on survival dominance (Theorem \ref{thm:lasbd2}). Association is typically not taken into account in models of reference-dependent preferences, e.g. the value function used in \citet{KosegiRabin06} is additive. Therefore, the necessary and sufficient conditions for dominance induced by this function are only a better distribution of gains/losses (in the sense of loss-aversion-sensitive dominance as in \citet{Firpoetal}) and a better distribution of income (in the sense of first-order stochastic dominance).  
Another set of results combines inequality aversion for incomes with inequality aversion for gains but inequality loving for losses. Theorem \ref{thm:iasd} provides a relevant dominance criterion, which combines second-order stochastic dominance with the criterion developed by \citet{LintonMaasoumiWhang05} for S-shaped value functions. The class of value functions should also be association-averse, cross-prudent and cross-temperate, or association-loving, cross-imprudent and cross-intemperate if one considers a parallel result that reverses the sign of higher-order derivatives of the value function (Theorem \ref{thm:iasd2}). 

All the criteria described above can be translated into sets of functional inequalities, in a manner analogous to the relationship between stochastic dominance and the functional inequality between distribution functions that is equivalent to dominance.  The dominance criteria all take into account more qualitative information on preferences, and as a result are not as simple as the typical stochastic dominance relationship, but tests can be designed that work the same way.  Our tests are similar in spirit to \citet{LintonSongWhang10} and involve the estimation of ``contact sets'', that is, (gain, income) pairs where the outcomes of two policies seem to be very similar.  The distribution of test statistics related to the inequalities can be expressed elegantly in the language of directionally-differentiable maps from distribution functions to test statistics, and allows us to borrow a special bootstrap method from \citet{FangSantos19} to conduct inference.  As shown in Section~\ref{sec:econometrics}, tests based on this bootstrap method are consistent against fixed alternatives and control size uniformly in the null region (the collection of probability measures where a dominance hypothesis is satisfied). 

The tests are illustrated using data from the experimental evaluation of a well-known welfare policy reform in the US in order to compare two policies that affect income distribution and generate gainers and losers. Specifically, we compare Aid to Families with Dependent Children (AFDC) with Jobs First (JF), the policy that replaced AFDC in Connecticut in the 1990s. Although Jobs First offered more generous income support than AFDC, it had a strict time limit beyond which no support was available. The evaluation randomly assigned households to either AFDC or Jobs First. These data were used by \citet{BitlerGelbachHoynes06}, who showed that, although the mean impact of Jobs First was positive, the policy created both winners and losers, meaning that its overall evaluation was not as straightforward as the mean impact suggested. \citet{BitlerGelbachHoynes06} used final incomes as an outcome. In a richer framework, where individuals care not only about their final income but also about the changes induced by the two policies, Jobs First is not the favored policy. Specifically, according to six dominance criteria developed in the paper, the hypothesis that JF is a dominant policy is always rejected, while the hypothesis that AFDC is a dominant policy is never rejected. More precisely, AFDC almost first-order dominates JF with respect to income changes, and our dominance criteria for income changes follow from this. We use these data only for illustrative purposes, namely to demonstrate the testing of the six criteria. Since the dominance criteria are linked to social welfare functions, one can conclude that JF is not a favored policy, and this conclusion holds across broad families of welfare functions. The joint evaluation further shows that JF’s advantage over AFDC is concentrated primarily among higher-income households with large gains, whereas AFDC provides a more favorable distribution elsewhere and, in particular, entails a lower risk of small losses across the board.




Our framework can be viewed as a framework of stochastic dominance \citep{Levy16}, and our results contribute directly to the stochastic dominance literature. The distributions being compared need not result from policy interventions; they may be any distributions, such as lotteries. Since our conditions for value functions can be interpreted as risk preferences, the results yield implementable criteria for evaluating lotteries in terms of these preferences. However, unlike in classic stochastic dominance, the dimensions are treated asymmetrically, with different conditions applying to each of them. 

Beyond the literatures on loss aversion in the behavioral economics, risk measurement and stochastic dominance, this paper also relates to the normative evaluation of tax policies. Traditionally, such evaluation is conducted by comparing the post-tax income distribution with the pre-tax distribution. However, this neglects the potential reranking of individuals induced by a tax reform \citep{AronsonJohnsonLambert94}. The issue of reranking is closely linked to horizontal equity, which requires that a fair tax system treat equals equally \citep{Musgrave59}. According to this principle, a tax reform should preserve the utility ranking of individuals \citep{Feldstein76, King83}. Such considerations naturally lead to a bidimensional framework, in which both the final distribution and the initial status quo distribution are taken into account \citep{AuerbachHassett02, Bourguignon11, Slesnick89}. For example, \citet{Bourguignon11} focuses on the joint distribution of status quo incomes and income changes, and shows that in this setting sequential stochastic dominance, as developed by \citet{AtkinsonBourguignon87}, can be applied to compare distributions. In that literature, differently than in this paper, the same dominance criteria applied to incomes are applied to income changes, with no attention to loss aversion or the S-shapedness of the value function that are the focus of this paper. Moreover, the status quo distribution plays a special role in that literature (a point that has been criticized, see e.g. \citep{Kaplow89, Kaplow95}). In our setting it is not necessary that the compared policies start from the same status quo. Here the distribution of gains and losses is important per se, rather than rerankings.       



The paper is organized as follows. The next three sections each develop the results linking qualitative features of utility functions to testable criteria based on joint distribution functions: Section \ref{sec:firstorder} (Theorem \ref{thm:lasbd} and \ref{thm:lasbd2}), Section \ref{sec:secondorder} (Theorem \ref{thm:iasd} and \ref{thm:iasd2}) and Section \ref{sec:liasdorder} (Theorem \ref{thm:liasd} and \ref{thm:liasd2}). We then relate the developed dominance conditions to functional inequalities and tests in Section~\ref{sec:empirical}, which also contains an empirical application. Section \ref{sec:conclusions} concludes. The appendix at the end of the paper contains proofs of the theorems.




\section{Loss Aversion Sensitive Bivariate dominance}\label{sec:firstorder}

Let $X$ be a random variable that denotes gains and losses with cumulative distribution function $F^1$ and density function $f^1$. Without loss of generality, let $\mathcal{X}=(-a_1, a_2) \subset \mathbb{R}$ denote the support of $X$.\footnote{For brevity, we often write $\int_{-\infty}^{\infty}$ instead of $\int_{-a_1}^{a_2}$.} Further, let $Z$ be a random variable that denotes outcomes in levels, with cumulative distribution function $F^2$ and density function $f^2$. Let $\mathcal{Z} = [0, a_3) \subset \mathbb{R}_{+}$ denote the support of $Z$. Finally, let $(X, Z)$ denote a random vector with joint cumulative distribution function $F$ and joint density function $f$. Let $\mathcal{X} \times \mathcal{Z}$ denote its support and let $\mathcal{F}$ denote the space of all bivariate distributions with support $\mathcal{X} \times \mathcal{Z}$. 

We define a bivariate social value function as follows.

\begin{definition}[Social Value Function (SVF)] \label{deff:svfbi}
    Let $W:\mathcal{F}\to\mathbb{R}$ denote the social value function
\begin{equation}
    W(F)=\int_{\mathbb{R}\times\mathbb{R}^+} v(x,z)\dd{F(x,z)},
\end{equation}
where $v: \mathcal{X} \times \mathcal{Z} \to\mathbb{R}$ is called a value function.
\end{definition}

\noindent Throughout the paper we consider various properties of the value function that define classes of SVF. We start with the following ones.

\begin{definition}[Loss aversion sensitive value function] \label{deff:lasbdproperty}
The value function $v:\mathcal{X} \times \mathcal{Z}\to\mathbb{R}$ is differentiable and satisfies:
\begin{itemize}
    \item Disutility of losses and utility of gains: $v(-x,z)\leq 0\leq v(x,z)$ for all $x>0, z$;
    \item Non-decreasing: $\frac{\partial}{\partial x} v(x,z)\geq 0$, $\frac{\partial}{\partial z} v(x,z)\geq 0$ for all $x,z$;
    \item Loss-averse: $\frac{\partial}{\partial x}v(-x,z)\geq \frac{\partial}{\partial x}v(x,z)$ for all $x>0,z$;
    \item Association averse (submodular): $\frac{\partial^2}{\partial x\partial z}v(x,z)\leq 0$ for all $x,z$.
\end{itemize}    
\end{definition}

\noindent The first three conditions regarding gains and losses in the Definition \ref{deff:lasbdproperty} are standard requirements in prospect theory \citep{TverskyKahneman91}: (i) losses hurt (bring negative utility) and gains bring utility, (ii) higher outcomes and higher gains (or smaller losses) are better, (iii) losses hurt more than gains of the same value. The fourth property is an aversion to positive associations between outcomes and gains/losses. That is, it is better to have more individuals with low outcomes but large gains and individuals with large outcomes but high losses than to have more individuals with both high outcomes and high gains or low outcomes and large losses. Association aversion is typically assumed in multivariate welfare and inequality measurement \citep{AtkinsonBourguignon82}, where two dimensions are typically two goods, e.g. income and life expectancy. It is understood as a preference for bringing individuals \textit{multidimensionally} closer together.

\begin{definition}[Loss Aversion Sensitive Bivariate Dominance] \label{deff:lasbd}
Let $(X_A, Z_A)$ and $(X_B, Z_B)$ have cumulative distribution functions respectively labeled $F_A, F_B\in\mathcal{F}$. If $W(F_A)\geq W(F_B)$ for all value functions $v$ that satisfy Definition \ref{deff:lasbdproperty}, we say that $F_A$ dominates $F_B$ in terms of Loss Aversion Sensitive Bivariate Dominance, or LASBD for short, and we write $F_A\succsim_{LASBD} F_B$.
\end{definition}

\begin{theorem}\label{thm:lasbd}
Suppose that $F_A, F_B\in\mathcal{F}$. The following are equivalent:
\begin{enumerate}
    \item $F_A\succsim_{LASBD} F_B$;
    \item For all $x\geq 0, z$, $F_A, F_B$ satisfy 
    \begin{equation}\label{eq:lasd}
        F^1_B(-x)-F^1_A(-x)\geq \max\{0,F^1_A(x)-F^1_B(x)\}
    \end{equation} 
    \begin{equation}\label{eq:fsd}
        F^2_A(z)\leq F^2_B(z)
    \end{equation}
     and for $x\neq a_2$ and $z \neq a_3$
    \begin{equation}\label{eq:fsbd}
        F_A(x,z)\leq F_B(x,z);
    \end{equation}
    \item For all $x\geq 0, z$, $F_A, F_B$ satisfy \eqref{eq:fsd} and \eqref{eq:fsbd} and the following conditions:
        \begin{equation}\label{eq:lasd1}
        F^1_A(-x)\leq F^1_B(-x)
    \end{equation}
        \begin{equation}\label{eq:lasd2}
        (1-F^1_A(x))-F^1_A(-x)\geq (1-F^1_B(x))-F^1_B(-x).
    \end{equation}
\end{enumerate}
\end{theorem}

\noindent Theorem \ref{thm:lasbd} is a natural extension of \citet{Firpoetal} to a bivariate setting. It states that ranking policy interventions (or distributions in general) over the class of social value functions, as in Definition \ref{deff:lasbdproperty}, is equivalent to the LASD dominance condition for gains and losses \eqref{eq:lasd} used in \citet{Firpoetal} (which, given their further results, is equivalent to \eqref{eq:lasd1} and \eqref{eq:lasd2}), first-order stochastic dominance for outcomes \eqref{eq:fsd}, and bivariate first-order stochastic dominance for the joint distribution of gains/losses and outcomes \eqref{eq:fsbd}. LASD is a consequence of loss aversion, and the rest is a consequence of the non-decreasingness of the value function; a well-known result is that for bivariate outcomes non-decreasingness and submodularity of the utility function is equivalent to bivariate first-order stochastic dominance \citep[see for example][]{Levy16}. Please note that for gains and losses, the condition does not follow from applying the joint condition to the marginal distribution, as in standard stochastic dominance. In this setting, preferences are asymmetric across the two dimensions, and consequently a weaker property than first-order dominance is required for income changes.    

A value function depending not only on the level of consumption but also on gains and losses is considered by \citet{KosegiRabin06}. They postulate a simple additive form, for which the dominance conditions degenerate to univariate conditions only, as the following corollary shows.  

\begin{corollary}\label{corr:additive}
    When the function $v$ is as in \citet{KosegiRabin06}: $v(x,z)=v_1(x)+v_2(z)$, then conditions \eqref{eq:lasd} and \eqref{eq:fsd} in Theorem \ref{thm:lasbd} are necessary and sufficient for $F_A \succsim_{LASBD} F_B$.
    \end{corollary}

While the first three properties in Definition \ref{deff:lasbdproperty} are building blocks of prospect theory, let us check what happens when we reverse the fourth property of submodularity of the value function. In a multivariate welfare and inequality measurement literature there are arguments for favoring higher association in some cases, for example, when goods are complements rather than substitutes \citep{BourguignonChakravarty03}. The appropriate definitions and dominance criteria are then the following.  

\begin{definition}[Loss Aversion Sensitive Bivariate Dominance 2] \label{deff:lasbd2}
Let $(X_A, Z_A)$ and $(X_B, Z_B)$ have cumulative distribution functions respectively labeled $F_A, F_B\in\mathcal{F}$. If $W(F_A)\geq W(F_B)$ for all value functions $v$ that satisfy Definition \ref{deff:lasbdproperty}, except that the last property of $v$ is association loving ($\frac{\partial^2}{\partial x\partial z}v(x,z)\geq 0$, that is, $v$ is supermodular) we say that $F_A$ dominates $F_B$ in terms of Loss Aversion Sensitive Bivariate Dominance 2, or LASBD2 for short, and we write $F_A\succsim_{LASBD2} F_B$.
\end{definition}

\begin{theorem}\label{thm:lasbd2}
Suppose that $F_A, F_B\in\mathcal{F}$. Let $K(x,z)=F^1(x)+F^2(z)-F(x,z)$. The following are equivalent:
\begin{enumerate}
    \item $F_A\succsim_{LASBD2} F_B$;
    \item For all $x\geq 0, z$, $F_A, F_B$ satisfy \eqref{eq:lasd}, \eqref{eq:fsd} and for $x\neq a_2$ and $z \neq a_3$
    \begin{equation}\label{eq:fsbd2}
        K_A(x,z)\leq K_B(x,z);
    \end{equation}
    \item For all $x\geq 0, z$, $F_A, F_B$ satisfy \eqref{eq:lasd1}, \eqref{eq:lasd2}, \eqref{eq:fsd} and \eqref{eq:fsbd2}.
\end{enumerate}
\end{theorem}


\noindent Compared to Theorem \ref{thm:lasbd}, Theorem \ref{thm:lasbd2} has a different condition \eqref{eq:fsbd2}. In \eqref{eq:fsbd} the integration is performed over rectangles $(-a_1, x) \times (0, z)$ and the condition is that the cumulative distribution function of the dominant distribution is everywhere less than or not greater than that of the dominated distribution. In contrast, \eqref{eq:fsbd2} is equivalent to the condition that the cumulative distribution of the dominant distribution over the rectangles $(x, a_2)\times (z, a_3)$ is everywhere greater than or equal to that of the dominated distribution. Therefore, dominance using $K_A, K_B$ is equivalent to dominance using a bivariate survival function. That is,
\[
K_A(x,z)\leq K_B(x,z) \Longleftrightarrow \int_x^{a_2} \int_z^{a_3} \dd F_A(x,z) \geq \int_x^{a_2} \int_z^{a_3} \dd F_B(x,z).
\]

\section{Inequality Aversion Sensitive Dominance}\label{sec:secondorder}

The LASBD dominance condition is of the `first order type'. It is similar to standard first-order stochastic dominance except that it accounts for loss aversion. Now we are interested in imposing more structure, which involves the consideration of inequalities induced by policy. Thus, the condition developed is of the `second order type'. In particular, the social planner will now be averse to inequality of outcomes and gains, which implies concavity of the value function. For losses, which are negative, the relevant requirement is the opposite, convexity. Overall, the value function is concave for outcomes and S-shaped for gains and losses. The latter is a standard second-order property of the value function in prospect theory \citep{LintonMaasoumiWhang05}.

\begin{definition}[Inequality aversion sensitive value function] \label{deff:iasdproperty}
The value function $v:\mathcal{X} \times \mathcal{Z}\to\mathbb{R}$ is differentiable and satisfies:
\begin{itemize}
    \item Disutility of losses and utility of gains: $v(-x,z)\leq 0\leq v(x,z)$ for all $x>0, z$;
    \item Non-decreasing: $\frac{\partial}{\partial x} v(x,z)\geq 0$, $\frac{\partial}{\partial z} v(x,z)\geq 0$ for all $x,z$;
    \item S-shaped in $x$: $\frac{\partial^2 v(x, z)}{\partial x^2} \leq 0$ and $\frac{\partial^2 v(-x, z)}{\partial x^2} \geq 0$ for all $x > 0$;
    \item Concave in $z$: $\frac{\partial^2 v(x, z)}{\partial z^2} \leq 0$;
    \item Association averse (submodular): $\frac{\partial^2}{\partial x\partial z}v(x,z)\leq 0$ for all $x,z$;
    \item Decreasingly submodular: $\frac{\partial^3}{\partial x^2\partial z}v(x,z)\geq 0$ and $\frac{\partial^3}{\partial x\partial z^2}v(x,z)\geq 0$;
    \item Cross-temperate: $\frac{\partial^4}{\partial x^2\partial z^2}v(x,z)\leq 0$.
\end{itemize}    
\end{definition}

\noindent The S-shape of the value function implies a preference for inequality in losses, but an aversion to inequality in gains. It is a preference for equal distribution of gains and unequal distribution of losses, e.g. for all losses to be concentrated in the smallest group of individuals. New properties are decreasing submodularity and cross-temperance. Decreasing submodularity means that not only is a  negative association between outcomes and gains/losses preferred (as in association aversion (submodularity)), but it is preferred especially at the bottom of the distribution, that is, among those who have high losses and low outcomes.  

Decreasing submodularity concerns the behaviour of the cross derivative of $x$ and $z$, but another way of looking at it is to consider the behaviour of the second order derivatives $\frac{\partial^2}{\partial x^2} v$ and $\frac{\partial^2}{\partial z^2} v$. Then the equivalent condition is that the value function is decreasingly concave for $z$ and $x$ and increasingly convex for $-x$ ($x>0$). That is, $\frac{\partial^3}{\partial x\partial z^2}v(x,z)\geq 0$ means that $\frac{\partial^2}{\partial z^2} v$ increases with $x$. Since we know that $\frac{\partial^2}{\partial z^2} v$ is negative (i.e. the fourth condition), this means that it is less negative at higher $x$, so the degree of concavity decreases with higher gains. In the same way we analyse $\frac{\partial^2}{\partial x^2} v$ for when $x<0$ and $x>0$. Decreasing concavity means that inequality is particularly hurtful at the bottom of the distribution. That is, the social planner prefers more equal outcomes among those who have lost the most than among those who have gained the most, and he/she also prefers more equal distribution of gains among the poorer (i.e. those with low incomes) than among the richer. For losses, on the other hand, the social planner prefers a more unequal distribution among the richer than among the poorer. The decreasing concavity of the value function corresponds to cross-prudence in risk measurement \citep{EeckhoudtReySchlesinger7, Jokung11}.  Cross-prudence is a preference for the disaggregation of two harms: a reduction in any attribute and the addition of zero mean risk to any attribute. A cross-prudent individual prefers to experience risk in one attribute if the value of the other attribute is higher rather than lower. For example, he prefers monetary risk when his health is better than when both monetary risk and health deterioration occur together. Replacing risk with inequality, more inequality in dimension is tolerated among those who have more of the other dimension. A similar criterion is often invoked in the context of socioeconomic inequalities in health and is said to reduce socioeconomic inequalities in health \citep[see e.g.][]{MakdissiYazbeck14}. In our context, as mentioned above, the social planner prefers, for example, higher outcome inequality among the winners than among the losers.

The final property, cross-temperance, relates to the fourth-degree derivative. In risk measurement, this means that the decision maker wants to disaggregate risks in both income and gains/losses. In our context, the social planner also prefers to disaggregate inequality in both income and gains/losses. Therefore, his preference for greater equality in one dimension increases with an increasing inequality in the other dimension. With cross-prudence the stronger the preference for equality in one dimension the lower the value of the other dimension; with cross-temperance the stronger the preference for equality in one dimension the less equal the other dimension is. 

\begin{definition}[Inequality Aversion Sensitive Dominance] \label{deff:iasd}
Let $(X_A, Z_A)$ and $(X_B, Z_B)$ have cumulative distribution functions respectively labeled $F_A, F_B\in\mathcal{F}$. If $W(F_A)\geq W(F_B)$ for all value functions $v$ that satisfy Definition \ref{deff:iasdproperty}, we say that $F_A$ dominates $F_B$ in terms of Inequality Aversion Sensitive Dominance, or IASD for short, and we write $F_A\succsim_{IASD} F_B$
\end{definition}

\begin{theorem}\label{thm:iasd}
Suppose that $F_A, F_B\in\mathcal{F}$. Let $H(x,z)=\int_{-\infty}^x\int_{0}^z F(t,s)\dd{s}\dd{t}$, $H^1(x)=\int_{-\infty}^{x} F^1(t) \dd{t}$, $S^1(x)=\int_x^{\infty} 1-F^1(t) \dd{t}$ and $H^2(z)=\int_{0}^z F^2(s)\dd{s}$. The following conditions are equivalent:
\begin{enumerate}
    \item $F_A\succsim_{IASD} F_B$;
    \item For all $x>0>y, z$, $F_A, F_B$ satisfy 
    \begin{equation}\label{eq:iasd1}
       S^1_A(x) - S^1_B(x) - \rbr{H^1_A (y) - H^1_B (y)} \leq S^1_A(0) - S^1_B(0) - \rbr{H^1_A (0) - H^1_B (0)},
    \end{equation} 
    \begin{equation}\label{eq:ssd}
        H^2_A(z) \leq H^2_B(z),
    \end{equation}
    and for $x \neq a_2$ and $z \neq a_3$ 
    \begin{equation}\label{eq:ssbd}
        H_A(x,z)\leq H_B(x,z);
    \end{equation}
    \item For all $x,z>0> y$, $F_A, F_B$ satisfy \eqref{eq:ssd}, \eqref{eq:ssbd} and
    \begin{equation}\label{eq:iasd}
       \int_{y}^x F^1_A(t)\dd{t}\leq \int_{y}^x F^1_B(t)\dd{t}.
    \end{equation}
\end{enumerate}
\end{theorem}

Theorem \ref{thm:iasd} states that the ranking of distributions induced by the class of social value functions with the properties described in Definition \ref{deff:iasdproperty} is equivalent to second-order stochastic dominance (\eqref{eq:ssd} and \eqref{eq:ssbd}), except that for gains and losses we obtain the dominance condition for S-shapedness of the value function \eqref{eq:iasd}. \citet{LintonMaasoumiWhang05} develop a test for this condition. As \citet[Theorem 4]{LevyWiener98} point out, the \eqref{eq:iasd} condition follows directly from considering gains (integral over $[0, x]$) and losses (integral over $[y, 0]$) separately, and assuming inequality aversion for the former and inequality loving for the latter. As Theorem \ref{thm:iasd} states it is equivalent to \eqref{eq:iasd1} for which we notice the following. Let $X^+, X^-$ denote, respectively, gains and losses (i.e. the positive and negative values of $X$). Then $S^1(0)=\mathrm{E}[X^+]$, that is, $S^1(0)$ is mean gain and $H^1(0)=\mathrm{E}[X^-]$ is mean loss. Furthermore $S^1(0)-H^1(0)=\mathrm{E}[X]$ is the mean of $X$. Thus \eqref{eq:iasd1} can be rewritten as 
\[
S^1_A(x) - S^1_B(x) - \rbr{H^1_A (y) - H^1_B (y)} \leq \mathrm{E}[X_A]-\mathrm{E}[X_B]
\]
for all $x>0>y$. Similarly to $S^1(0)$ and $H^1(0)$, $S^1(x)$ can be interpreted as the average gain \textit{above} $x$ and $H^1(y)$ as the average loss \textit{below} $y$. Thus, condition \eqref{eq:iasd1} states that for $F_A$ to dominate $F_B$ for gains and losses, it has to be that, for all $x,y$, the difference between the distributions' differences in average gain above $x$ (i.e. $S^1_A(x) - S^1_B(x)$) and in average loss below $y$ (i.e. $H^1_A (y) - H^1_B (y)$), is smaller than the difference in distributions' mean gains and losses (i.e. $\mathrm{E}[X_A]-\mathrm{E}[X_B]$). Moreover, when $\mathrm{E}[X_A^-] = \mathrm{E}[X_B^-]$ i.e. mean loss is the same for $F_A$ and $F_B$, $\int_{-\infty}^{y} F^1_{A}(t) \dd t \geq \int_{-\infty}^{y} F^1_{B}(t) \dd t$ has to hold in the losses region for all $y$. This is a typical second order stochastic dominance condition but applied to the space of losses. For losses then, the distribution that yields lower welfare ($F_{B}$) second-order stochasticaly dominates the distribution that brings higher welfare ($F_{A}$). So the standard condition is reversed, which is not surprising given that S-shapedness means that losses are evaluated using a convex, not a concave function as is typically the case.  


We have a parallel class of value functions with some of the preferences changed. 

\begin{definition}[Inequality Aversion Sensitive Dominance 2] \label{deff:iasdproperty2}
Let $(X_A, Z_A)$ and $(X_B, Z_B)$ have cumulative distribution functions respectively labeled $F_A, F_B\in\mathcal{F}$. If $W(F_A)\geq W(F_B)$ for all value functions $v$ that satisfy Definition \ref{deff:iasdproperty}, with exception that they are
\begin{itemize}
     \item Association loving (supermodular): $\frac{\partial^2}{\partial x\partial z}v(x,z)\geq 0$ for all $x,z$;
    \item Decreasingly supermodular: $\frac{\partial^3}{\partial x^2\partial z}v(x,z)\leq 0$ and $\frac{\partial^3}{\partial x\partial z^2}v(x,z)\leq 0$;
    \item Cross-intemperate (second-degree supermodular)$\frac{\partial^4}{\partial x^2\partial z^2}v(x,z)\geq 0$;
\end{itemize} we say that $F_A$ dominates $F_B$ in terms of Inequality Aversion Sensitive Dominance 2, or IASD2 for short, and we write $F_A\succsim_{IASD2} F_B$.
\end{definition}

Compared to Definition \ref{deff:iasdproperty}, value functions in Definition \ref{deff:iasdproperty2} favor the association between gains(losses) and outcomes. That is, it is preferable to have more individuals in a society with high (resp. low) incomes and high gains (resp. high losses) than to have those for whom one dimension has higher value and the other has lower value.  Decreasing supermodularity is equivalent to increasing concavity for $z$ and $x$ and increasing convexity for $-x$, where $x>0$. Increasing concavity comes from the second order derivatives of $z$ and $x>0$, both of which are negative, being even more negative (i.e. increasingly concave) with $x$ and $z$ respectively. Cross-intemperance comes from the fact that the social planner prefers to aggregate inequalities in both dimensions and therefore his preference for equality in one dimension decreases with the degree of inequality in the other dimension.  

\begin{theorem}\label{thm:iasd2}
Suppose that $F_A, F_B\in\mathcal{F}$. Let $L(x,z)=\int_{-\infty}^x\int_{0}^z K(t,s)\dd{s}\dd{t}$. 
The following conditions are equivalent:
\begin{enumerate}
    \item $F_A\succsim_{IASD2} F_B$;
    \item For all $x>0>y, z$, $F_A, F_B$ satisfy \eqref{eq:iasd1} and \eqref{eq:ssd}
    and for $x\neq a_2$ and $z \neq a_3$ 
    \begin{equation}\label{eq:ssbd2}
        L_A(x,z)\leq L_B(x,z).
    \end{equation}
\end{enumerate}
\end{theorem}
Since the properties of $v$ for $x$ and $z$ are the same as in Definition \ref{deff:iasdproperty}, conditions \eqref{eq:iasd1} and \eqref{eq:ssd} from Theorem \ref{thm:iasd} are also necessary and sufficient in Theorem \ref{thm:iasd2}. The change is in the condition for the joint \eqref{eq:ssbd2}, which is a second-order condition involving $K$ (Theorem \ref{thm:lasbd2}). 

\section{Loss and Inequality Aversion Sensitive Dominance} \label{sec:liasdorder}

Our most interesting condition combines loss aversion in gains and losses with inequality aversion in outcomes, as they are the most often postulated preferences when it comes to gains/losses and outcomes. Furthermore, the value function is averse to the positive association of gains/losses and outcomes, particularly so at the bottom of the distribution. There is also a preference for the disaggregation of inequalities in gains/losses and outcomes.  

\begin{definition}[Loss and inequality aversion sensitive value function] \label{deff:liasdproperty}
The value function $v:\mathcal{X} \times \mathcal{Z}\to\mathbb{R}$ is differentiable and satisfies:
\begin{itemize}
    \item Disutility of losses and utility of gains: $v(-x,z)\leq 0\leq v(x,z)$ for all $x>0, z$;
    \item Non-decreasing: $\frac{\partial}{\partial x} v(x,z)\geq 0$, $\frac{\partial}{\partial z} v(x,z)\geq 0$ for all $x,z$;
    \item Loss-averse in $x$: $\frac{\partial}{\partial x}v(-x,z)\geq \frac{\partial}{\partial x}v(x,z)$ for all $x>0,z$;
    \item Concave in $z$: $\frac{\partial^2 v(x, z)}{\partial z^2} \leq 0$;
    \item Association averse (submodular): $\frac{\partial^2}{\partial x\partial z}v(x,z)\leq 0$ for all $x,z$;
    \item Decreasingly submodular: $\frac{\partial^3}{\partial^2 x\partial z}v(x,z)\geq 0$ and $\frac{\partial^3}{\partial x\partial^2 z}v(x,z)\geq 0$;
    \item Cross-temperate (second-degree submodular)$\frac{\partial^4}{\partial^2 x\partial^2 z}v(x,z)\leq 0$.
\end{itemize}    
\end{definition}

As before, loss and inequality aversion can be formalized as a dominance concept.

\begin{definition}[Loss and Inequality Aversion Sensitive Dominance] \label{deff:liasd}
Let $(X_A, Z_A)$ and $(X_B, Z_B)$ have cumulative distribution functions respectively labeled $F_A, F_B\in\mathcal{F}$. If $W(F_A)\geq W(F_B)$ for all value functions $v$ that satisfy Definition \ref{deff:liasdproperty}, we say that $F_A$ dominates $F_B$ in terms of Loss and Inequality Aversion Sensitive Dominance, or LIASD for short, and we write $F_A\succsim_{LIASD} F_B$
\end{definition}

As with the previous classes of value functions, dominance in this class of value functions can be translated to a set of conditions on the distributions of the results of two policies.

\begin{theorem}\label{thm:liasd}
Suppose that $F_A, F_B\in\mathcal{F}$. The following conditions are equivalent:
\begin{enumerate}
    \item $F_A\succsim_{LIASD} F_B$;
    \item For all $x, z$, $F_A, F_B$ satisfy \eqref{eq:lasd} and \eqref{eq:ssd} and for $x\neq a_2$ and $z \neq a_3$ they satisfy \eqref{eq:ssbd}.
\end{enumerate}
\end{theorem}

According to the previous results, Theorem \ref{thm:liasd} follows quite naturally. The dominance condition prescribed in Theorem \ref{thm:liasd} is the sum of the conditions for loss aversion in gains and losses and second-order stochastic dominance \eqref{eq:ssd} and \eqref{eq:ssbd}. That is, to rank policy interventions in a way that takes into account loss aversion in gains and losses and inequality aversion in final outcomes, it is necessary to check LASD dominance for gains/losses, second order stochastic dominance for outcomes, and bivariate second order stochastic dominance for the joint distribution. 

We have a parallel class of social value function that preserves loss aversion and inequality aversion, but has different conditions for cross derivatives. 

\begin{definition}[Loss and Inequality Aversion Sensitive Dominance 2] \label{deff:liasd2}
Let $(X_A, Z_A)$ and $(X_B, Z_B)$ have cumulative distribution functions respectively labeled $F_A, F_B\in\mathcal{F}$. If $W(F_A)\geq W(F_B)$ for all value functions $v$ that satisfy Definition \ref{deff:liasdproperty}, with exception that it is
\begin{itemize}
     \item Association loving (supermodular): $\frac{\partial^2}{\partial x\partial z}v(x,z)\geq 0$ for all $x,z$;
    \item Decreasingly supermodular: $\frac{\partial^3}{\partial x^2\partial z}v(x,z)\leq 0$ and $\frac{\partial^3}{\partial x\partial z^2}v(x,z)\leq 0$;
    \item Cross-intemperate (second-degree supermodular)$\frac{\partial^4}{\partial x^2\partial z^2}v(x,z)\geq 0$;
\end{itemize}
we say that $F_A$ dominates $F_B$ in terms of Loss and Inequality Aversion Sensitive Dominance 2, or LIASD2 for short, and we write $F_A\succsim_{LIASD2} F_B$.
\end{definition}

For this class of functions we have the following result.

\begin{theorem}\label{thm:liasd2}
Suppose that $F_A, F_B\in\mathcal{F}$. The following conditions are equivalent:
\begin{enumerate}
    \item $F_A\succsim_{LIASD2} F_B$;
    \item For all $x, z$, $F_A, F_B$ satisfy \eqref{eq:lasd} and \eqref{eq:ssd} and for $x\neq a_2$ and $z \neq a_3$ they satisfy \eqref{eq:ssbd2}. 
\end{enumerate}
\end{theorem}

\noindent Naturally, Theorem \ref{thm:liasd2} is a combination of previously used conditions, \eqref{eq:lasd} as in Theorem \ref{thm:lasbd} and \eqref{eq:ssd} and \eqref{eq:ssbd2} as in Theorem \ref{thm:iasd2}, as it combines loss aversion in $x$ with inequality aversion in $z$ and favors association in the joint distribution.

\section{Statistical inference} \label{sec:econometrics}
We can test the systems of functional inequalities implied by the various dominance criteria.  In this section we explain how to conduct these tests.

Each of the dominance concepts defined above is re-expressed using several conditions that describe the relationship between observable features of the dominant and dominated distributions.  In particular, all the conditions can be written as functional inequalities, checking for rejection of a dominance hypothesis by checking the sign of the corresponding function.

For example, the first LASBD definition is equivalent to conditions on the joint and marginal distribution functions of distributions $A$ and $B$ as described in Theorem~\ref{thm:lasbd}.  To conduct inference about the dominance of distribution $A$ over $B$, we convert $H_0^1: F_A \succsim_{LASBD} F_B$ into an equivalent hypothesis on a set of functional inequalities provided by (in the case of the LASBD criterion) displays~\eqref{eq:fsd}-\eqref{eq:lasd2}.

We search for deviations from the null by rearranging the conditions into functions that should be everywhere nonpositive, and search for arguments where that appears to be violated significantly.  In the case of the first LASBD condition, we can define the test function
\begin{equation*}
    g(x, z) = g(F_A, F_B)(x, z) = g^{LASBD}(F_A)(x, z) - g^{LASBD}(F_B)(x, z)
\end{equation*}
where
\begin{equation*}
    g^{LASBD}(F)(x, z) = \begin{bmatrix}  F^2(z) \\ 
    F(-x,z) \\
    F(x,z) \\
    F^1(-x) \\
    F^1(x) + F^1(-x) \end{bmatrix},
\end{equation*}
and the null hypothesis $H_0^1$ can be rewritten
\begin{equation*}
    H_0^2: g(x, z) \leq \mathbf{0}_5 \quad \forall\, x, z \geq 0.
\end{equation*}
For reference, functions analogous to $g^{LASBD}$ for all the dominance concepts discussed here are collected in the appendix.

We need a way to measure deviations from the hypothesized inequalities, that is, to find $(x, z)$ pairs where it appears that at least one coordinate of the test function is positive. These functions can all be estimated in a straightforward way using plug-in estimates, that is, the empirical (joint and marginal) distribution functions from each observed sample.  We define $\hat{g} = g(\hat{F}_A, \hat{F}_B)$, where $(\hat{F}_A, \hat{F}_B)$ are the empirical distribution functions for $(F_A, F_B)$.  Under the null hypothesis, letting $[x]_+ = \max\{x, 0\}$ (applied coordinate-wise to vectors) and $\|\cdot\|$ be the $L_2$ norm, $\|[g]_+\| = 0$.  Therefore we should have $\|[\hat{g}]_+\| \approx 0$,
where the test statistic should only be positive due to random sampling.

The continuous mapping theorem implies that the test statistic $T_n = \|[\hat{g}]_+\|$ converges to zero in probability to zero under the null.  However, the asymptotic distribution of $T_n$ is intractable.  Usually, one would turn to the bootstrap in this situation, but the pointwise map $x \mapsto [x]_+$ in the definition of the statistic complicates the distribution.  We find in convenient to interpret the map $(F_A, F_B) \mapsto \|[g(F_A, F_B)]_+\|$ as a Hadamard directionally differentiable transformation of a pair of distributions into a real-valued statistic (Hadamard directional differentiability of similar maps was discussed extensively in \citet{FGP23, Firpoetal}).  This characterization allows for inference with a modified bootstrap procedure, as described in \citet{FangSantos19}.  There is other research that might apply to the testing of this problem: \citet{LeeSongWhang18} propose a general method for testing functional inequalities, and an alternative bootstrap is described by \citet{HongLi20}.  The method described below is tailored specifically to these tests.

The tests of all the dominance hypotheses work in the same way and can be described in general.  We assume that the null hypothesis has been translated into a multivariate functional inequality in $(x, z)$, and that the test function $g$ is nonpositive under the null that $A$ dominates $B$ in the chosen sense.  We call its plug-in estimate $\hat{g}$ and its estimate using a bootstrapped sample is labeled $g^\ast$.  Then a hypothesis test is conducted this way:

\begin{enumerate}
    \item Estimate $\hat{g}$ and $T_n = \|[\hat{g}]_+\|$ using plug-in estimates of the distribution functions.
    \item Use $\hat{g}$ to estimate the contact set, that is, the collection of $(x, z)$ such that $g(x, z) = 0$.  Call the contact set estimator function $\chi_{c_n}(x, z) = I(|\hat{g}(x, z)| \leq c_n)$.  Use $c_n \searrow 0$ such that $c_n \sqrt{n} \rightarrow \infty$ (in our empirical example, we use $c_n = 4 \log \log n / \sqrt{n}$, as suggested by simulation evidence in \citet{LintonSongWhang10}).
    \item For each iteration $r = 1, \ldots, R$ of the bootstrap, resample the data with replacement and calculate $g_r^\ast$ and $T_r^\ast = \|[(g_r^\ast - \hat{g}) \cdot \chi_{c_n}]_+\|$.
    \item Use the reference distribution $\{T_r^\ast\}_{r=1}^R$ for inference: for example, the bootstrap p-value is, for arbitrarily small $\eta > 0$,
    \begin{equation*}
        p^\ast = \frac{1}{R} \sum_{r=1}^R I(T_r^\ast + \eta > T_n).
    \end{equation*}
\end{enumerate}

The parts of this algorithm that are not typical of all bootstrap inference techniques are the parts involving contact sets in steps 2 through 4.  The reasons behind steps 2 and 3 will be seen in Theorem~\ref{thm:consistent} below, and the reason for the $\eta$ in step 4 will become apparent after the statement of Theorem~\ref{thm:size}.  

The formal results stating the consistency of this bootstrap procedure for inference with the loss and inequality averse dominance criteria rely on two minimal regularity assumptions that describe the sample data we assume to be observable.

\begin{enumerate}[label=\textbf{A\arabic*}]
  \item \label{assumptionA_first} The data are continuously distributed with marginal distribution functions $F_A$ and $F_B$ respectively.  The observations $\{X_{Ai}\}_{i=1}^{n_A}$ and $\{X_{Bi}\}_{i=1}^{n_B}$ are i.i.d. samples of size $n_A$ and $n_B$ and the samples are independent of each other.
  \item \label{assumptionA_last} Define $n = n_A + n_B$. $n_A$ and $n_B$ increase such that $n_k / n \rightarrow \lambda_k$ as $n_A, n_B \rightarrow \infty$, where $0 < \lambda_k < 1$ for $k \in \{A, B\}$.
\end{enumerate}

Under the minimal assumptions above, we can show that the bootstrap distribution is a consistent estimator of the limiting distribution of the test statistics for all the dominance concepts.  In the following two statements, we use the following notation.  We let $BL_1$ be the space of real-valued Lipschitz functions that are bounded by 1, which is a space of functions typically used to make statements about the weak convergence of a sequence of random variables to its distributional limit.  The operators $\mathrm{P}\{\cdot\}$ and $\mathrm{E}[\cdot]$ denote the probability measure and expected value using the population distribution, while $\mathrm{P}^\ast\{\cdot\}$ and $\mathrm{E}^\ast[\cdot]$ refer to the counterparts using the distribution of the bootstrapped data conditional on the observed sample.  The equality $X_n = o_P(1)$ implies that the sequence $X_n$ converges in probability to zero as $n$ diverges.  Although many dominance concepts were defined and discussed above, the statistical analysis of tests used for all of the concepts is qualitatively identical, so we refer to all of them in the same way.  They all use a sample test function $g(\hat{F}_A, \hat{F}_B)$ to learn about the population test function $g = g(F_A, F_B)$, where the individual coordinates of $g$ may change with each dominance concept, and for each concept there is a set of distributions $\mathcal{F}_0 \subset \mathcal{F}$ that satisfy the null hypothesis, and make it so that $(F_A, F_B) \in \mathcal{F}_0$ implies $T = \|[g]_+\| = 0$.  We will refer to all test functions and all null collections as $g$ and $\mathcal{F}_0$ in the theorems below, although they change with the particular notion of dominance.

\begin{theorem} \label{thm:consistent}
Let $T_n$ be any of the statistics described above for testing, that is, $T_n = \|[g(\hat{F}_A, \hat{F}_B)]_+\|$ for any of the $g$ described in the appendix.  Under conditions~\ref{assumptionA_first} and~\ref{assumptionA_last}, $T_n$ converges weakly to $T = \|[g]_+\|$ in the sense that there exists a random variable $\mathcal{T}$ such that
\begin{equation*}
    \sup_{f \in BL_1} \left\lvert \mathrm{E}\left[ f\left( \sqrt{n}(T_n - T) \right) \right] - \mathrm{E}\left[ f\left( \mathcal{T} \right) \right] \right\rvert = o_P(1).
\end{equation*}
Similarly, if $T_n^\ast = \|[g(F_A^\ast, F_B^\ast) - g(\hat{F}_A, \hat{F}_B)]_+ \cdot \chi_{c_n}\|$ denotes the analogous test function evaluated with bootstrap empirical distributions and sample empirical distributions, we have
\begin{equation*}
    \sup_{f \in BL_1} \left\lvert \mathrm{E}^\ast\left[ f\left( \sqrt{n}T_n^\ast \right) \right] - \mathrm{E}\left[ f\left( \mathcal{T} \right) \right] \right\rvert = o_P(1).
\end{equation*}
\end{theorem}

The second part of Theorem~\ref{thm:consistent} has one unexpected part, which is the form that the bootstrap analog $T_n^\ast$ takes.  In particular, it is not perfectly analogous to the form that weak convergence takes with the sample data because of the presence of the contact set indicator.  Because of the positive part map that lies in the definition of the statistics, they are fundamentally less regular than other more conventional statistics and a special bootstrap needs to be devised to ensure bootstrap consistency.  This bootstrap technique relies on \citet{FangSantos19} and is justified in another way in \citet{LintonSongWhang10}.

The previous theorem asserted the consistency of the resampling plan.  The next theorem adds to that description.  It specifies the size of testing procedures used to infer dominance as described in Sections~\ref{sec:firstorder}, \ref{sec:secondorder} and~\ref{sec:liasdorder}.  To do so, we introduce a sequence of local alternative distributions $(F_{An}, F_{Bn})$ and assume that they satisfy a kind of mean-square convergence condition to the null $(F_A, F_B)$: assume that for $F_n = F_{An}$ or $F_{Bn}$, and $F = F_A$ or $F_B$, there is some square integrable $h$ such that
\begin{equation} \label{eq:local}
    \lim_{n \rightarrow \infty} \int \left( \sqrt{n}(\sqrt{\mathrm{d} F_n} - \sqrt{\mathrm{d} F}) - \frac{h}{2} \sqrt{\mathrm{d} F} \right)^2 \rightarrow 0.
\end{equation}
This form of alternative is used in empirical process theory \citep[\S 3.11.1]{vanderVaartWellner23} to discuss distributions that converge to the null at precisely the right rate to find nontrivial limit results.

\begin{theorem} \label{thm:size}
Assume conditions~\ref{assumptionA_first} and~\ref{assumptionA_last} are satisfied. Assume that $(F_A, F_B) \in \mathcal{F}_0$, that is, that $g = g(F_A, F_B)$ satisfies $T = \|[g]_+\| = 0$ for one of the $g$ described in the appendix, and let $T_n = \|[g(\hat{F}_A, \hat{F}_B)]_+\|$.  Letting $q(1 - \alpha)$ denote the $(1-\alpha)$-th quantile of the asymptotic distribution of $\sqrt{n}T_n$, suppose that $q(1 - \alpha) > 0$.  Suppose that a local sequence of probability distributions $P_n = (F_{An}, F_{Bn})$ satisfies~\eqref{eq:local} and $\|[g(F_{An}, F_{Bn})]_+\| = 0$ for all $n$.  Finally, let $q_n^\ast(1-\alpha)$ denote the $(1-\alpha)$-th quantile of $T_n^\ast = \|[g(F_A^\ast, F_B^\ast) - g(\hat{F}_A, \hat{F}_B)]_+ \cdot \chi_{c_n}\|$. Then $\limsup_{n \rightarrow \infty} \mathrm{P}_n \left\{ \sqrt{n}T_n^\ast > q_n^\ast(1 - \alpha) \right\} \leq \alpha$. This holds with equality when $P_n = P_0$ for all $n$.
\end{theorem}

This result implies that the bootstrap test's size is controlled asymptotically by the intended/nominal test size for the distribution $P_0$ and all local alternatives that respect the null hypothesis of dominance of $A$ over $B$.  One could in theory compute the power of tests for alternatives that violate the null hypothesis, but this is complicated by the unique features of each testing criterion and no general (yet practical) statements can be made about test power under local alternatives.  The regularity condition that the $(1-\alpha)$-th quantile of the asymptotic distribution of the test statistic must be positive made in the previous theorem may seem innocuous.  However, it has practical implications.  If, for example, $F_A$ and $F_B$ are such that we are ``far'' from rejecting the null hypothesis, it is possible that $T_n = 0$ and $T_r^\ast = 0$ for all the bootstrap repetitions.  In this case, the naive bootstrap p-value $p^\circ = \sum_r I(T_r^\ast > T_n) / R = 0$.  However, the distribution is degenerate here and this seemingly-low p-value does not indicate that the observed test statistic lies in an extreme region of the reference distribution.  To address this, we suggest using the modified bootstrap critical value akin to that proposed in \citet{AndrewsShi13}, namely $p^\ast = \sum_r I(T_r^\ast + \eta > T_n) / R$, where $\eta > 0$ is an arbitrarily small constant like $\eta = 10^{-6}$.  This has the effect of breaking ties due to degeneracy when they happen, and has no practical effect otherwise.

\section{Empirical illustration: Welfare reform in Connecticut} \label{sec:empirical}

In this section we illustrate the comparisons discussed above using household data from an experiment conducted by the U.S. state of Connecticut in 1996. This data has been discussed at length before \citep{BitlerGelbachHoynes06, Firpoetal} so we only briefly describe the main features of the two policies and the patterns that emerge in the sample. 

The treatments in this experiment are both programs that provided income support to low-income families with dependent children.  The preexisting Aid to Families with Dependent Children (AFDC) program was replaced in the 1990s with a different program called Temporary Assistance for Needy Families (TANF). The specific TANF program that was administered by Connecticut was called Jobs First (JF) and had a much different structure than AFDC: it included more generous income support than the AFDC program had, but that support came with with a strict time limit. We label the two treatments as AFDC and JF benefit structures. The state was interested in comparing program outcomes and so it experimentally assigned one of AFDC or JF benefit structures to a sample of 4803 households (we observe 2396 households under treatment JF and 2407 households under treatment AFDC). For each household, we observe quarterly income before the treatment was assigned, during the experiment and also after the shorter JF time limit had been reached. Outcomes are measured in the natural logarithm of average household income over post-experiment periods (quarter years), and gains/losses are simply the log post-experiment average minus the log pre-experiment average. \citet{BitlerGelbachHoynes06}, using average and quantile treatment effects in levels, found that JF made the majority of households better off, but also had significant drawbacks for some households after the time limit had been reached.  We will make different comparisons using the loss- and inequality-averse criteria developed above, focusing on household income data.

\begin{figure}[ht]
\includegraphics[width=\columnwidth]{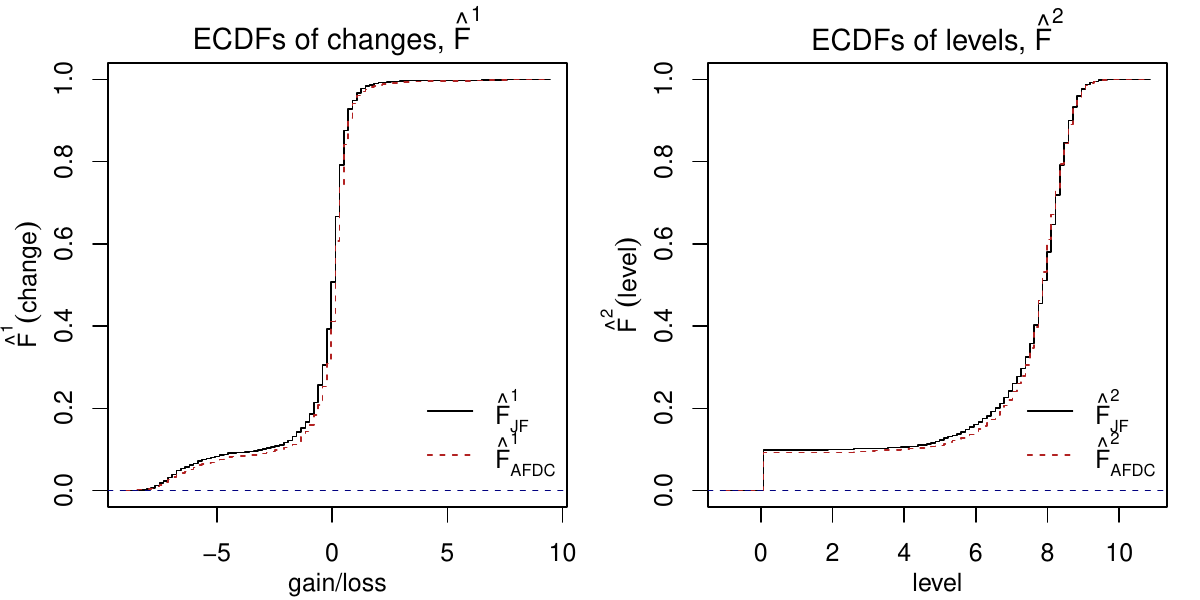}
\caption{Marginal empirical CDFs of income distributions under $JF$ and $AFDC$ benefit structures, which indicate the distributions of gains/losses and levels respectively. The marginal AFDC gain/loss distribution stochastically dominates the JF gain/loss distribution at first order. In levels, the marginal ECDFs cross.}
\label{fig:margins}
\end{figure}

Let us first take a look at a comparison of marginal and joint income distributions under both policies. They are plotted in Figure~\ref{fig:margins}. We can see that the empirical CDF (ECDF) of income \emph{changes} looks better under the AFDC policy over the entire support of the change distributions. The dominance appears quite strong, namely, the AFDC curve seems to stochastically dominate the JF curve at first order. However, when looking at post-experiment household income \emph{levels}, the policies are not clearly ordered. The levels ECDFs cross, with the marginal AFDC distribution function below that of JF for lower-income households. Above level $z = 7$, the ECDFs cross. This level corresponds to around 2300 US dollars on average quarterly. It is difficult to see in the plots, but above that level of income, the JF and AFDC level distribution functions cross several times. The first-order dominance of AFDC over JF in changes implies that of the loss averse and `second order type' comparisons of the marginal distribution of changes in the test functions, and most likely leads to the pattern of rejections and non-rejections shown in the table of tests below. Therefore, these empirical results should be regarded primarily as illustrations of the methods and their associated testing procedures.

\begin{figure}[ht]
\includegraphics[width=\columnwidth]{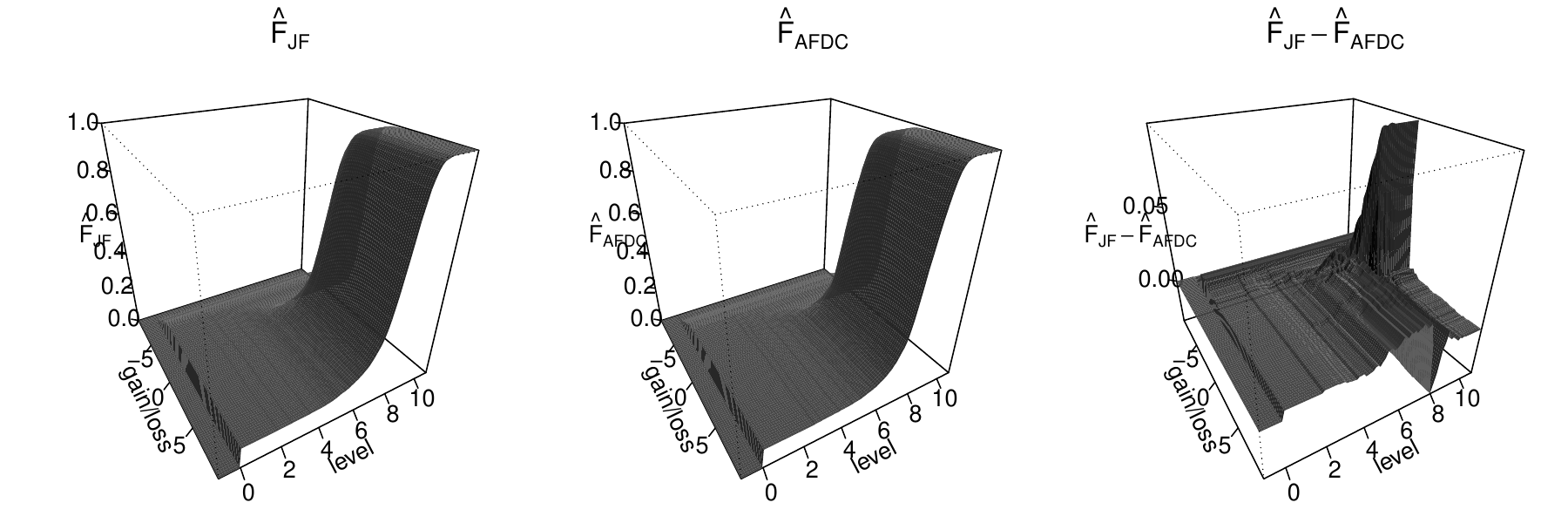}
\caption{Joint empirical CDFs $\hat{F}_{JF}$ and $\hat{F}_{AFDC}$ and their difference, since the ECDFs look very similar.  See the main text for a longer discussion of the spike and dip at the right side of the third panel.}
\label{fig:joint}
\end{figure}

The left two panels of Figure~\ref{fig:joint} show the joint ECDFs of both benefit distributions. The third panel shows the difference between the joint ECDFs, which is not readily apparent in the left two panels. For reference, the ``front'' corner of all three plots can be used to see where zero is on the vertical axis.  The difference in the third panel is calculated such that positive parts indicate that the JF distribution function lies above the AFDC distribution function. The third panel most clearly shows information that cannot be gained by only comparing marginal ECDFs. Broadly speaking, the AFDC structure has a joint distribution function that lies below the JF joint distribution function everywhere except for a dramatic reversal for high-income households. For these households, JF produces a higher likelihood of large gains than AFDC—this is where JF’s advantages primarily emerge, namely in the high-income, high-gain range. Consequently, households that fare better under JF compared with AFDC are found predominantly among those with higher incomes. Conversely, the positive spike in the region of small gains and losses indicates that high-income households also face a higher risk of losses under JF than under AFDC. 

We now turn to statistical tests to assess the significance of these observations. We ran tests to check whether JF dominates AFDC.  These are shown in the left half of Table~\ref{tab_pvals}.  We also ran tests to check whether AFDC dominates JF, and those results are presented in the right half of the table.  For an exact test statistic, one would need to evaluate the empirical process upon which the test statistics rely at each sample observation, but the combined sample size is prohibitively large. Instead, empirical processes were evaluated on a grid of points in the $(x, z)$ space that used 100 grid points for gains \& losses, and 50 points for levels, and test statistics were computed as functionals of that approximated process.  Informal experimentation with denser grids resulted in extremely similar results.  999 bootstrap repetitions were used to estimate a reference distribution.  Once again, this number could be increased but the results are qualitatively similar.

Several hypothesis testing results are presented in Table~\ref{tab_pvals}. Using any of the six dominance concepts developed above, we reject the hypothesis that the JF benefits structure would be preferred by households. On the other hand, we are unable to reject the hypothesis that AFDC benefits would be preferred by households. There are some variations.  For example, the first-order-type LASBD test is less decisive than the second-order-type IASD and LIASD tests.  However, the basic result is unchanged across all the dominance concepts.

\begin{table}[!tbp]
\begin{center}
\begin{tabular}{lrrcrr}
\hline\hline
\multicolumn{1}{l}{\bfseries }&\multicolumn{2}{c}{\bfseries $H_0:$ JF $\succsim$ AFDC}&\multicolumn{1}{c}{\bfseries }&\multicolumn{2}{c}{\bfseries $H_0:$ AFDC $\succsim$ JF}\tabularnewline
\cline{2-3} \cline{5-6}
\multicolumn{1}{l}{}&\multicolumn{1}{c}{statistic}&\multicolumn{1}{c}{p-value}&\multicolumn{1}{c}{}&\multicolumn{1}{c}{statistic}&\multicolumn{1}{c}{p-value}\tabularnewline
\hline
LASBD&$0.23$&$0.02$&&$0.05$&$0.54$\tabularnewline
LASBD2&$0.27$&$0.00$&&$0.06$&$0.48$\tabularnewline
IASD&$7.53$&$0.00$&&$0.01$&$0.81$\tabularnewline
IASD2&$5.66$&$0.00$&&$0.57$&$0.44$\tabularnewline
LIASD&$7.40$&$0.00$&&$0.01$&$0.78$\tabularnewline
LIASD2&$5.49$&$0.00$&&$0.57$&$0.42$\tabularnewline
\hline
\end{tabular}
\caption{Tests of the hypotheses that either JF benefits dominate AFDC benefits or that AFDC benefits dominate JF benefits.  In all cases, the JF benefits appear to violate the hypothesis that they would be preferred by households.  On the other hand, we cannot reject the hypothesis that AFDC benefits dominate JF benefits under any of the dominance concepts.  999 bootstrap repetitions were used to generate reference distributions for all tests, and functions were evaluated on evenly-spaced grids of 100 points for gains/losses and 50 points for incomes in levels.\label{tab_pvals}}\end{center}
\end{table}

The hypothesis that JF dominates AFDC is rejected using any dominance concept described in this paper. On the other hand, tests of the dominance of AFDC over JF fail to reject in all cases.  Assuming that households' behavior is well described by any of the sets of qualitative properties given in the previous sections, there is strong evidence that the AFDC benefit structure is socially preferred comparing to the JF. It appears as though JF carries a greater likelihood of small losses to household income than AFDC, suggesting that many households had been supplementing their earnings with program support and, when JF assistance ended, their incomes declined.

In the following subsections, we show coordinate processes used in all the tests shown in the left half of Table~\ref{tab_pvals}.  All the functions are found by rearranging the functions in the corresponding inequalities shown earlier in the text so that the function corresponding to the JF benefit structure has the AFDC function subtracted from it~--- for example, in Figure~\ref{fig:lasbd_common} below, the left plot shows $\hat{F}_{JF}^2(z) - \hat{F}_{AFDC}^2(z)$ over all levels $z$, used to test whether the inequality~\eqref{eq:fsd} holds (squared values of the positive part of this function are combined in an integral with similar quantities using other coordinate functions to calculate a test statistic).

\subsection{Loss aversion sensitive comparison}

We illustrate the way that the distributions are compared using the LASBD and LASBD2 concepts. Both the LASBD and LASBD2 concepts use changes in household income from before the program started, when all households were using AFDC benefits, to after the program ended when households subjected to the JF treatment saw their benefits end.  This risk, that a household could potentially have a lower income if the JF benefits end and there is no other sizable source of income, is the risk of the JF program that would be of primary concern to a household considering the two policies.

\begin{figure}[H]
\includegraphics[width=\columnwidth]{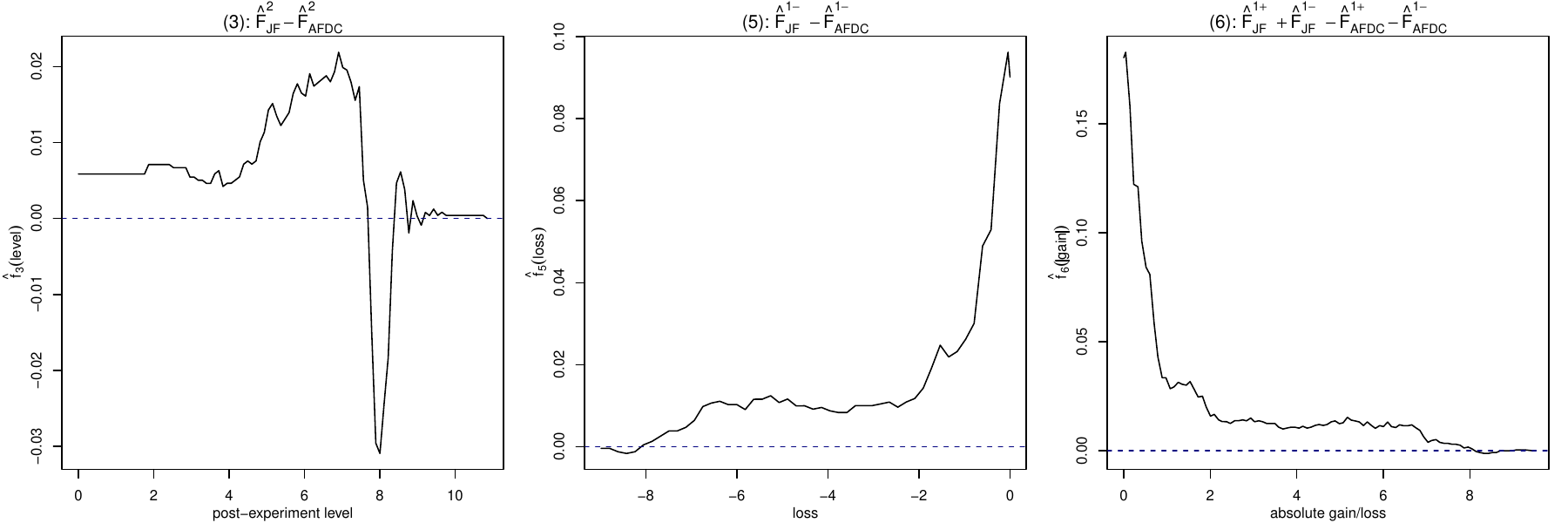}
\caption{The component functions common to the LASBD and LASBD2 notions used for testing either $H_0: F_{JF} \succsim_{LASBD} F_{AFDC}$ or $H_0: F_{JF} \succsim_{LASBD2} F_{AFDC}$.  The numbers above each panel correspond to the numbered displays in the text.  $F_k^{1+}$ is shorthand for the distribution function $k$ for gains/losses evaluated for gains and $F_k^{1-}$ is the same function evaluated for losses.}
\label{fig:lasbd_common}
\end{figure}

\begin{figure}[H]
\includegraphics[width=\textwidth]{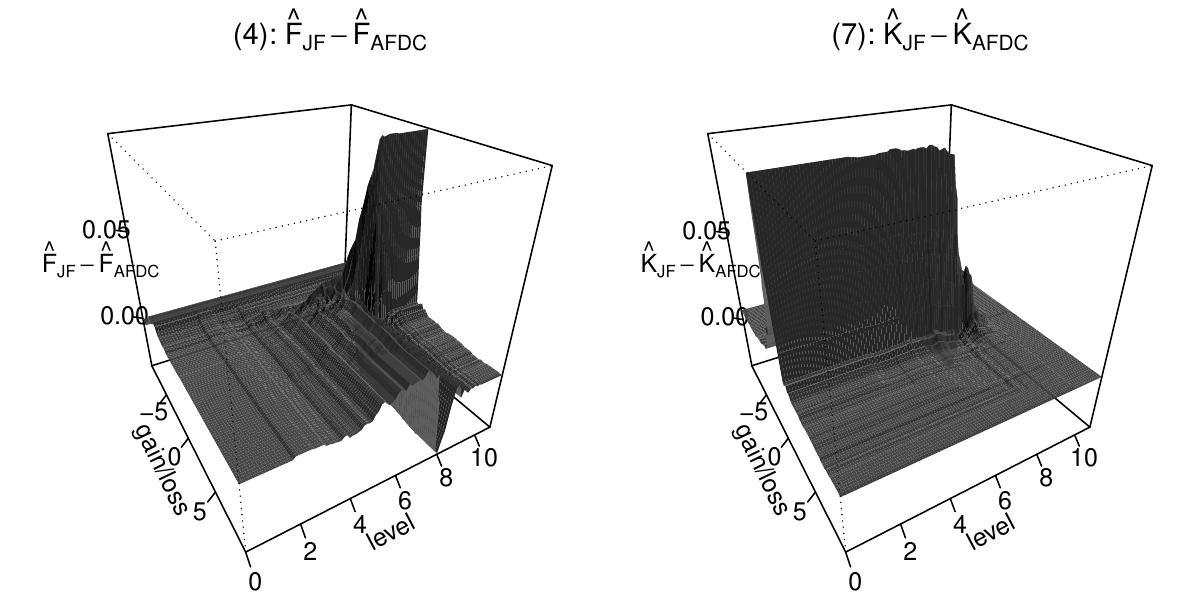}
\caption{The component functions that differ between the LASBD and LASBD2 concepts.  The left panel is analogous to display~\eqref{eq:fsbd} and shows $\hat{F}_{JF}(x, z) - \hat{F}_{AFDC}(x, z)$ while the right panel is analogous to display~\eqref{eq:fsbd2} and shows $\hat{K}_{JF}(x, z) - \hat{K}_{AFDC}(x, z)$, both evaluated using the empirical distribution estimates of $F_{JF}$ and $F_{AFDC}$.  For reference in these 3-dimensional plots, the functions are zero at the ``corners'' of the plots, where $x$ and $z$ reach either of their extremes, indicating that the central portions are positive.} 
\label{fig:lasbd_different}
\end{figure}

Figure~\ref{fig:lasbd_common} and Figure~\ref{fig:lasbd_different} show all the component functions that go into a test of either of the nulls $F_{JF} \succsim_{LASBD} F_{AFDC}$ or $F_{JF} \succsim_{LASBD2} F_{AFDC}$.  In testing, these two concepts have three common component functions, shown in Figure~\ref{fig:lasbd_common}, and differ in one component function, which are contrasted in Figure~\ref{fig:lasbd_different}.  In the paragraphs to follow, we describe what causes positive values seen in the component functions, which would indicate a violation of either $F_{JF} \succsim_{LASBD} F_{AFDC}$ or $F_{JF} \succsim_{LASBD2} F_{AFDC}$.  These positive values are reduced to one-number summaries that are in the top two entries of the leftmost column of Table~\ref{tab_pvals}.

All three functions in Figure~\ref{fig:lasbd_common} are positive for some arguments $(x, z)$.  The reasons are different for each panel of the figure.  The left panel tracks income levels after the JF time limit.  It indicates that $F_{JF}^2(z) \geq F_{AFDC}^2(z)$, roughly speaking for $z$ up to the middle of the income distribution~--- in other words, the post-experiment incomes are more favorable for low-income households under the AFDC benefit structure.  The relationship between the distributions changes at $\exp(7.75) \approx \$2300$, at which point $F_{JF}$ goes below $F_{AFDC}$ and then the functions stay close for larger quarterly incomes.  The central panel of Figure~\ref{fig:lasbd_common} reveals that (using~\eqref{eq:lasd1}) $F_{AFDC}^1 \leq F_{JF}^1$ for almost all levels of loss, but $F_{JF}$ represents a much higher chance of experiencing a small loss as compared to $F_{AFDC}$, leading to the large values on the right-hand side of the plot.  The right panel of Figure~\ref{fig:lasbd_common} is positive for small values of $|x|$ representing absolute gains and losses as in~\eqref{eq:lasd2}, and is positive because $F_{AFDC}^1$ dominates $F_{JF}^1$ for small changes especially.

As mentioned above, the LASBD and LASBD2 concepts share three coordinate functions.  They differ in one coordinate, which is illustrated in Figure~\ref{fig:lasbd_different}.  $F_A \succsim_{LASBD} F_B$ requires that $A$ dominates $B$ in the bivariate distribution function, while $F_A \succsim_{LASBD2} F_B$ requires it dominate in $K$ functions (which are the probabilities that a gain/level pair exceed $(x, z)$). Both functions are positive for some $(x, z)$, suggesting a rejection of the null hypotheses $F_{JF} \succsim_{LASBD} F_{AFDC}$ or $F_{JF} \succsim_{LASBD2} F_{AFDC}$.  For LASBD, this is because $F_{JF}$ does not dominate $F_{AFDC}$ for moderately large income levels and all changes above small losses.  That is, the probability of moderately high post-experiment incomes are relatively high and coupled with (usually) some gain in income under AFDC, while for JF that probability is not as high.  The LASBD2 concept shows nearly the same information~--- under $F_{JF}$, households have a lower probability of having high incomes and small gains.

\subsection{Inequality aversion sensitive comparison}

\begin{figure}[H]
\includegraphics[width=\textwidth]{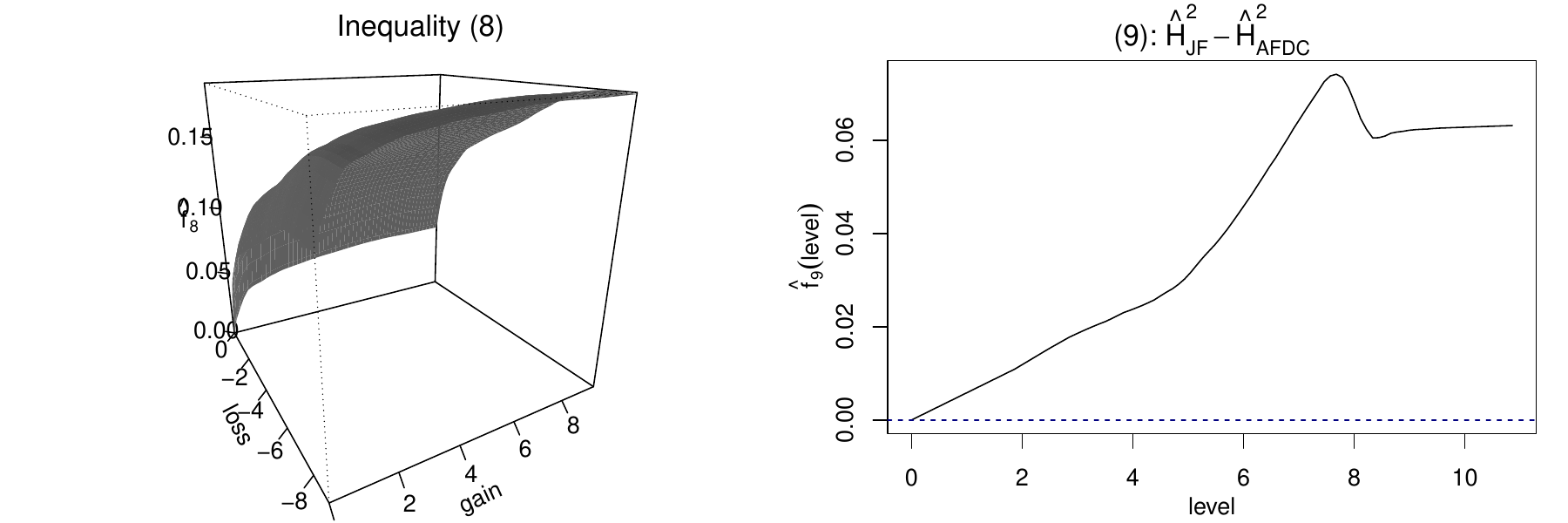}
\caption{These functions are common to the IASD and IASD2 comparison between JF and AFDC.  Because they remain above zero everywhere, they suggest a contradiction of the hypotheses $F_{JF} \succsim_{IASD} F_{AFDC}$ or $F_{JF} \succsim_{IASD2} F_{AFDC}$.}
\label{fig:iasd_common}
\end{figure}

We can make a similar comparison between the functions that go into the IASD and IASD2 criteria.  There are two functions that IASD and IASD2 have in common, and are shown in Figure~\ref{fig:iasd_common}.

Both of the functions are exactly zero at the origin, but become positive when evaluating the function anywhere else.  Because they are positive, they suggest a violation of the notion that the hypothesis that the JF benefit structure would be preferred by households to the AFDC benefit structure in an IASD sense, that is, they indicate there is evidence against the hypothesis $F_{JF} \succsim_{IASD} F_{AFDC}$.

\begin{figure}[H]
\includegraphics[width=\textwidth]{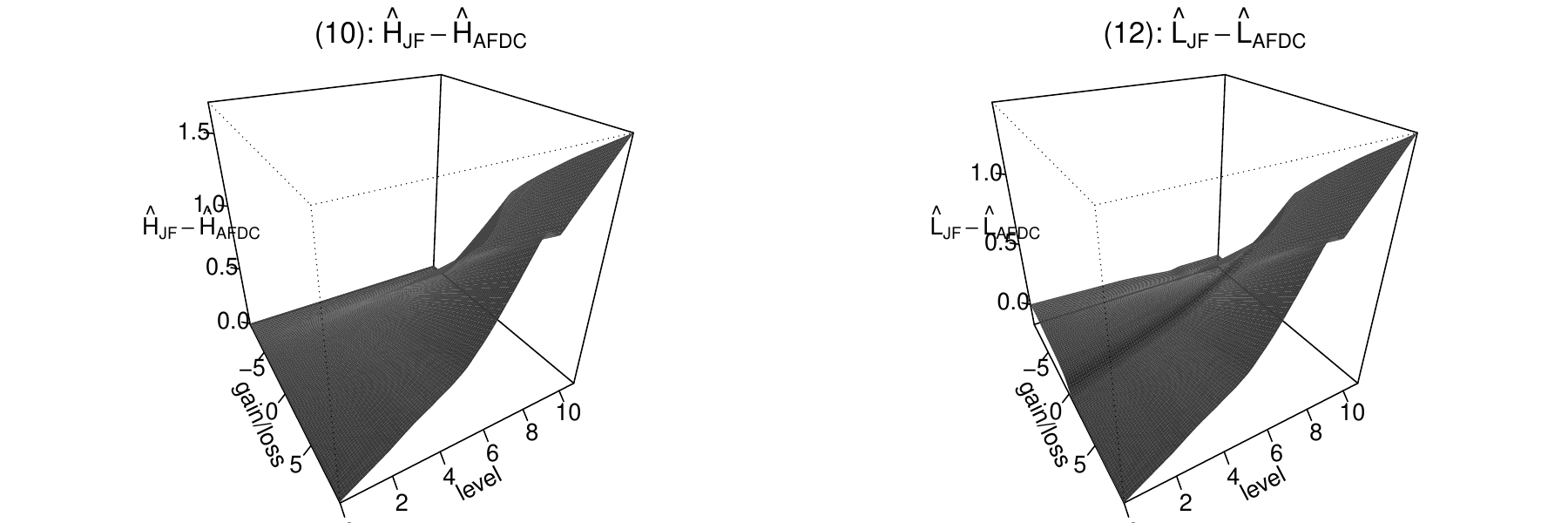}
\caption{These functions are different between the IASD and IASD2 comparison of JF and AFDC.  The IASD concept uses function~\eqref{eq:ssbd}, pictured in the left panel, and the IASD2 concept uses function~\eqref{eq:ssbd2}, in the right panel.  Both functions remain above zero over their support, with the exception of the right panel (function~\eqref{eq:ssbd2}) for low-income households that experience a large gain over the course of the experiment.  The large positive values these functions take suggest a violation of the hypotheses that JF dominates AFDC using either the IASD or IASD2 criterion.}
\label{fig:iasd_different}
\end{figure}

The difference between the IASD and IASD2 criteria are that IASD uses equation~\eqref{eq:ssbd} while IASD2 uses~\eqref{eq:ssbd2} for comparison.  These two functions are shown in Figure~\ref{fig:iasd_different}. At the left-most corner of each panel of Figure 6, the functions are equal to zero, and they are increasing as gain/loss move away from their lower limits.  Once again, these functions are constructed so that significantly positive values suggest a rejection of the hypothesis $F_{JF} \succsim_{IASD} F_{AFDC}$.

\subsection{Loss- and inequality aversion sensitive comparison}

The LIASD comparison of these two programs is similar, and indeed, it recombines some of the coordinate functions already pictured in LASBD and IASD comparisons.  To test the hypothesis $F_{JF} \succsim_{LIASD} F_{AFDC}$, one would look for positive values of the functions~\eqref{eq:lasd}, \eqref{eq:ssd} and~\eqref{eq:ssbd}.  Equation~\eqref{eq:lasd} is easily checked by using equations~\eqref{eq:lasd1} and~\eqref{eq:lasd2}, which are shown in the center and right panels of Figure~\ref{fig:lasbd_common}.  The other two coordinate functions were used in the IASD comparison, and their plots can be found in the right panel of Figure~\ref{fig:iasd_common} and the left panel of Figure~\ref{fig:iasd_different}.  The large positive values in all these plots suggest a rejection of the hypothesis, but of course the second-to-last test statistic and its p-value in the left half of Table~\ref{tab_pvals} is used to formally test their joint statistical significance.

The story is similar for a test of the hypothesis $F_{JF} \succsim_{LIASD2} F_{AFDC}$, which also uses functions~\eqref{eq:lasd} and~\eqref{eq:ssd} and were discussed in the previous paragraph.  The way that the LIASD2 comparison differs is in its use of~\eqref{eq:ssbd2}, and the difference $L_{JF} - L_{AFDC}$ can be seen in the right panel of Figure~\ref{fig:iasd_different}.  These functions also indicate a rejection of the LIASD2 dominance of JF over AFDC, as seen formally in the final row of the left half of Table~\ref{tab_pvals}.

\section{Conclusions}\label{sec:conclusions}

We propose a variety of bivariate stochastic dominance criteria that may be used to rank prospective policies based on experimental data on both the gains and losses that agents experience and their post-policy outcome in levels. They can also be used to rank any bivariate distributions of absolute outcomes and outcome changes, for example, lotteries. We extend existing univariate approaches so that they incorporate two central empirical regularities that are well-grounded in the literature: individuals dislike losses more than they value equivalent gains, and they are sensitive to the distribution of absolute incomes. The dominance criteria remain non-parametric, which ensures that the rankings of distributions that are obtained are robust to the choice of specific functional forms of value functions. Furthermore, the criteria can be translated into sets of functional inequalities, and testing methods are designed to check whether one policy/distribution is preferred to another. 

In this paper we underscore the importance of jointly considering changes in income and the distribution of final incomes. From a policy point of view it is important to know where gains and losses are concentrated along the income dimension. Furthermore, policies that appear attractive when assessed solely on gains and losses may be much less appealing once their distributional consequences are incorporated, and vice versa. The bivariate dominance framework reduces the risk of adopting policies that are ``extreme'' in a unidimensional sense, instead favoring those that perform robustly well across both key dimensions of individual preferences.

While this is one of the first papers to incorporate loss aversion into welfare analysis, it is important to acknowledge the ongoing debate on this issue in the behavioral economics literature. As already mentioned, we rely on this literature by using an extended utility function whose properties are well established in this literature. It is interesting to note that this literature has almost entirely avoided welfare analysis because it is unclear whether reference dependence represents a bias on the part of decision-makers or non-standard but normative preferences \citep{ReckSeibold23}. Or, as \citet{ODonoghueSprenger18} put it: ``Perhaps first and foremost is the question of whether gain-loss utility should be given normative weight i.e., whether we should assume that the same preferences that rationalize behavior should also be used for welfare analysis.'' In other words, while empirical evidence shows that individuals maximize reference-dependent preferences when making decisions, the question is whether the social planner should use these same preferences as an input in the social welfare maximization problem. If the social planner believes that such preferences represent errors and distort behavior from what would be individually optimal, then paternalistically the social planner should only take into account the ``correct'' preferences. That question is certainly relevant to our framework. 

We agree with the argument put forward in \citep{TverskyKahneman91}, namely, that decision makers need a good criterion for evaluating policy options, and it is hard to argue that actual experience of the consequences of a policy can be completely discarded as such a criterion. People do evaluate their situations in relation to the reference point, and they do feel an asymmetry between pain and pleasure. Apart from the numerous citations above, this is also well supported by neuroeconomic research \citep{Dhami16} and is known to influence the behavior of humans and primates in general. This leads \citet{ReesJones24} to state ``Modern economists have been wary of taking this type of paternalistic stance... who is the researcher to say, confidently, that they know what is best for others?''. Therefore, very recently, a welfare analysis literature is emerging in behavioral economics \citep{GoldinReck22, ReckSeibold23} that includes the loss/gain component and uses an extended utility function. It is, however, different formally than our framework as it includes individual maximizing behavior as a component. Our welfare analysis, as noted, is grounded in the traditional social welfare function approach.            

From a broader perspective, this work contributes to the emerging welfare-economic literature that integrates behavioral features—such as loss aversion—into social evaluation without imposing a particular functional specification. It provides a set of implementable tools for researchers and policymakers who wish to respect empirically observed preference patterns while maintaining the robustness and transparency of dominance-based methods. Future research could apply these criteria in other contexts where reference-dependent preferences are relevant, explore their interaction with dynamic considerations such as persistence of losses or intertemporal inequality, and develop social value indices that are characterized by further properties and useful in cases in which the dominance criteria do not produce conclusive results.

\newpage
\appendix
\noindent \begin{Large}\textbf{Appendix}\end{Large} 
\vspace{-0.5cm}

\section*{Functions used in testing}
The hypotheses to be tested in the main text can be translated into tests of functional inequalities of one type.  For each dominance concept, we write a test function $g = g^k(F_A) - g^k(F_B)$ where $k$ is a stand-in for a concept's label.  Recall that the CDF $F$ is bivariate, and the marginal CDFs corresponding to $F$ are labeled $F^1$ and $F^2$ (for gains/losses and levels respectively).  These functions $g$ can be evaluated over $x, z \geq 0$~--- gains and losses are represented by positive or negative values of $x$ respectively and are treated separately, while levels are represented by the argument $z$ (recall it is assumed that $z \in [0, a_3) \subset \R_+$).  In the definitions below, we use the following functions, which are also stated in the theorems defining the functions used for testing in the main text:

\begin{align*}
    H^1(x) &= \int_{-\infty}^x F^1(t) \dd t, \quad H^2(z) = \int_0^z F^2(t) \dd t & S^1(x) &= \int_x^\infty 1 - F^1(t) \dd t \\
    K(x, z) &= F^1(x) + F^2(z) - F(x, z) & L(x, z) &= \int_{-\infty}^x \int_0^z K(t, s) \dd s \dd t.
\end{align*}

Then we define the following functions for testing (for IASD/IASD2, we need $x_1, x_2 \geq 0$):

\footnotesize
\begin{align*}
    g^{LASBD}(F)(x, z) &= \begin{bmatrix} F^2(z) \\ F(-x, z) \\ F(x, z) \\ F^1(-x) \\ F^1(x) + F^1(-x) \end{bmatrix} & g^{LASBD2}(F)(x, z) &= \begin{bmatrix} F^2(z) \\ K(-x, z) \\ K(x, z) \\ F^1(-x) \\ F^1(x) + F^1(x) \end{bmatrix}, \\
    g^{IASD}(F)(x_1, x_2, z) &= \begin{bmatrix} H^2(z) \\ H(-x_1, z) \\ H(x_1, z) \\ S^1(x_1) - H^1(-x_2) - S^1(0) + H^1(0) \end{bmatrix} & g^{IASD2}(F)(x_1, x_2, z) &= \begin{bmatrix} H^2(z) \\ L(-x_1, z) \\ L(x_1, z) \\ S^1(x_1) - H^1(-x_2) - S^1(0) + H^1(0) \end{bmatrix}, \\
    g^{LIASD}(F)(x, z) &= \begin{bmatrix} H^2(z) \\ H(-x, z) \\ H(x, z) \\ F^1(-x) \\ F^1(x) + F^1(-x) \end{bmatrix} & g^{LIASD2}(F)(x, z) &= \begin{bmatrix} H^2(z) \\ L(-x, z) \\ L(x, z) \\ F^1(-x) \\ F^1(x) + F^1(-x) \end{bmatrix}.
\end{align*}
\normalsize

\section*{Proof of main results}
Let us start by formulating the following lemma, which will be useful later in the proofs. 

\begin{lemma}\label{lem:svf2bi}
\begin{multline}\label{eq:svf2bi}
W(F)=-\int_{-\infty}^{0}\frac{\partial}{\partial x}v(x,a_3) F^1(x)\dd{x}+\int_{0}^{\infty}\frac{\partial}{\partial x}v(x,a_3) (1-F^1(x))\dd{x}-\int_0^\infty \frac{\partial}{\partial z}v(a_2,z) F^2(z)\dd{z} \\ +\int_{-\infty}^{\infty}\int_0^\infty \frac{\partial^2}{\partial x\partial z}v(x,z)F(x,z)\dd{z}\dd{x}
\end{multline}
\end{lemma}
\begin{proof}[Proof of Lemma \ref{lem:svf2bi}]
\[
W(F)=\int_{\mathbb{R}\times\mathbb{R}^+} v(x,z)\dd{F(x,z)}=\int_{-\infty}^{\infty}\int_0^\infty v(x,z)f(x,z)\dd{z}\dd{x}
\]
\[
=\int_{-\infty}^{\infty}\sbr{v(x,z)\int_0^z f(x,t)\dd{t}|_0^{a_3}-\int_0^\infty \frac{\partial}{\partial z}v(x,z)\int_0^z f(x,t)\dd{t} \dd{z}}\dd{x}
\]
\[
=\int_{-\infty}^{\infty}v(x,a_3) f^1(x)\dd{x}-\int_{-\infty}^{\infty}\int_0^\infty \frac{\partial}{\partial z}v(x,z)\int_0^z f(x,t)\dd{t} \dd{z}\dd{x}.
\]

We will now expand these two terms. We have 
\[
\int_{-\infty}^{\infty}v(x,a_3) f^1(x)\dd{x}=\int_{-\infty}^{0}v(x,a_3) f^1(x)\dd{x}+\int_{0}^{\infty}v(x,a_3) f^1(x)\dd{x}.
\]
Starting with the first term and applying integration by parts we get
\[
\int_{-\infty}^{0}v(x,a_3) f^1(x)\dd{x}=v(x,a_3) F^1(x)|_{-a_1}^0-\int_{-\infty}^{0}\frac{\partial}{\partial x}v(x,a_3) F^1(x)\dd{x}=-\int_{-\infty}^{0}\frac{\partial}{\partial x}v(x,a_3) F^1(x)\dd{x}
\]
and also
\[
\int_{0}^{\infty}v(x,a_3) f^1(x)\dd{x}=-\int_{0}^{\infty}v(x,a_3) \dd{\rbr{1-F^1(x)}}=-v(x,a_3)(1-F^1(x))|_0^{a_2} +\int_{0}^{\infty}\frac{\partial}{\partial x}v(x,a_3) (1-F^1(x))\dd{x}=\]\[=\int_{0}^{\infty}\frac{\partial}{\partial x}v(x,a_3) (1-F^1(x))\dd{x}.
\]

Expanding the second term, we have
\[
\int_{-\infty}^{\infty}\int_0^\infty \frac{\partial}{\partial z}v(x,z)\int_0^t f(x,t)\dd{t} \dd{z}\dd{x} = \int_0^\infty\frac{\partial}{\partial z}v(x,z) F(x,z)\dd{z}|_{-a_1}^{a_2}-\int_{-\infty}^{\infty}\int_0^\infty \frac{\partial^2}{\partial x\partial z}v(x,z)F(x,t)\dd{z}\dd{x}=
\]
\[
\int_0^\infty \frac{\partial}{\partial z}v(a_2,z) F^2(z)\dd{z}-\int_{-\infty}^{\infty}\int_0^\infty \frac{\partial^2}{\partial x\partial z}v(x,z)F(x,z)\dd{z}\dd{x}
\]

Finally, we obtain
\[
W(F)=-\int_{-\infty}^{0}\frac{\partial}{\partial x}v(x,a_3) F^1(x)\dd{x}+\int_{0}^{\infty}\frac{\partial}{\partial x}v(x,a_3) (1-F^1(x))\dd{x}\]\[-\int_0^\infty \frac{\partial}{\partial z}v(a_2,z) F^2(z)\dd{z}+\int_{-\infty}^{\infty}\int_0^\infty \frac{\partial^2}{\partial x\partial z}v(x,z)F(x,z)\dd{z}\dd{x}
\]
\end{proof}

\begin{proof}[Proof of Theorem \ref{thm:lasbd}]
Notice that $\eqref{eq:lasd}$ is equivalent to both $\eqref{eq:lasd1}$ and $\eqref{eq:lasd2}$ and we will use latter conditions. Let $\Delta W = W(F_A)-W(F_B)$ and similarly for cumulative distribution functions, $\Delta F = F_A - F_B$. Using Lemma \ref{lem:svf2bi} we have that
\[
\Delta W =-\int_{-\infty}^{0}\frac{\partial}{\partial x}v(x,a_3) \Delta F^1(x)\dd{x}-\int_{0}^{\infty}\frac{\partial}{\partial x}v(x,a_3)\Delta F^1(x)\dd{x}\]\[-\int_0^\infty \frac{\partial}{\partial z}v(a_2,z) \Delta F^2(z)\dd{z}+\int_{-\infty}^{\infty}\int_0^\infty \frac{\partial^2}{\partial x\partial z}v(x,z)\Delta F(x,z)\dd{z}\dd{x}=
\]
 \[
=-\int_0^{\infty}\frac{\partial}{\partial x}v(-x,a_3) \Delta F^1(-x)\dd{x}-\int_{0}^{\infty}\frac{\partial}{\partial x}v(x,a_3)\Delta F^1(x)\dd{x}\]\[-\int_0^\infty \frac{\partial}{\partial z}v(a_2,z) \Delta F^2(z)\dd{z}+\int_{-\infty}^{\infty}\int_0^\infty \frac{\partial^2}{\partial x\partial z}v(x,z)\Delta F(x,z)\dd{z}\dd{x} \geq 0
\]   
    
\noindent Adding and subtracting $ \int_0^{\infty} \frac{\partial}{\partial x}v(x,a_3)\Delta F^1(-x)\dd{x}$ we get
    \begin{multline}\label{eq:t1bi}
     \Delta W =   \int_0^{\infty}\frac{\partial}{\partial x}(v(x,a_3)-v(-x,a_3)) \Delta F^1(-x)\dd{x}-\int_{0}^{\infty}\frac{\partial}{\partial x}v(x,a_3)(\Delta F^1(x)+\Delta F^1(-x))\dd{x}\\-\int_0^\infty \frac{\partial}{\partial z}v(a_2,z) \Delta F^2(z)\dd{z}+\int_{-\infty}^{\infty}\int_0^\infty \frac{\partial^2}{\partial x\partial z}v(x,z)\Delta F(x,z)\dd{z}\dd{x} \geq 0. 
    \end{multline}

\noindent Utilizing the assumptions of loss aversion, non-decreasingness and submodularity given in Definition \ref{deff:lasbdproperty}, \eqref{eq:fsd} and \eqref{eq:fsbd}, \eqref{eq:lasd1} and \eqref{eq:lasd2} (or, equivalently \eqref{eq:lasd}) are sufficient for \eqref{eq:t1bi} to hold. 


We now show that these conditions are also necessary by means of a contradiction. For the first two integrals in \eqref{eq:t1bi} the procedure is a modification of \citet{Firpoetal} to two dimensions. Starting with \eqref{eq:lasd1}, from the fact that the distribution function is right continuous, there is a neighborhood $(a, b)$, $0<a<b$, such that for all $x\in(a,b)$, $F^1_A(-x)-F^1_B(-x)>0$ (i.e. $\Delta F^1 (-x) >0$). Now consider the value function
\[
v_1(x, z)=\begin{cases}
    a-b & x\leq -b\\
    x+a & x\in (-b,-a)\\
    0 & x\geq -a
\end{cases}
\]
Note that $v_1(x, z)$ satisfies Definition \ref{deff:lasbdproperty}.\footnote{The function is not differentiable at the boundaries of the three areas, however, we can always find a differentiable function which is as close to $v_1$ as one wishes too, e.g. for arbitrarily small $\epsilon>0$ we consider $\frac{1}{4\varepsilon}(x+b+\varepsilon)^2 +a-b$ when $x\in(-b-\varepsilon,-b+\varepsilon)$ that then joins areas $x\leq -b-\varepsilon$ and $(-b+\varepsilon,-a-\varepsilon)$. The same applies to the other value functions in the proofs.} In particular, it is non-decreasing with respect to $z$, because its derivative with respect to $z$ is $0$. It is submodular in a trivial way, that is, cross-derivative is $0$. It is also loss averse because, for $x\geq0$. $\frac{\partial v(x, z)}{\partial x} =0$, while for $x<0$ the respective derivative is $1$ when $x \in (-b,-a)$. Therefore
$\int_0^{\infty} \frac{\partial}{\partial x}(v(x, a_3)-v(-x, a_3))\Delta F^1 (-x)\dd{x} <0 $ while the rest of integrals in \eqref{eq:t1bi} are $0$, which contradicts \eqref{eq:t1bi}.
Condition \eqref{eq:lasd2} can be proven similarly. Assume that there exists a neighborhood $(a,b)$, $0<a<b$ such that for all $x\in(a,b)$
$(1-F^1_A(x))-F^1_A(-x)< (1-F^1_B(x))-F^1_B(-x)$ (i.e. $\Delta F^1(x)+\Delta F^1(-x) > 0$). Take $v_2(x, z)=-v_1(-x, z)$ for $x>0$ and $v_2(x, z)=v_1(x, z)$ for $x<0$, then  $-\int_{0}^{\infty}\frac{\partial}{\partial x}v(x,a_3)(\Delta F^1(x)+\Delta F^1(-x))\dd{x} < 0$ while the rest of integrals in \eqref{eq:t1bi} are $0$, which contradicts \eqref{eq:t1bi}.

Now we proceed in a similar fashion with the third integral in \eqref{eq:t1bi} and a contradiction to \eqref{eq:fsd}. Assume that there exists some $z>0$ such that $F^2_A(z)-F^2_B(z)>0$ (i.e. $\Delta F^2(z)>0$). From the fact that the distribution function is right continuous, it follows that there is a neighborhood $(c, d)$, $0<c<d$, such that for all $z\in(c,d)$, $F^2_A(z)-F^2_B(z)>0$. For $x\geq0$, consider the value function 
\[
v_3(x, z)=\begin{cases}
    0 & z\leq c\\
    z-c & z\in (c,d)\\
    d-c & z\geq d
\end{cases}
\]
and for $x < 0$ we put $bx, b>0$. Thus, $v_3$ fulfills Definition \ref{deff:lasbdproperty}. In particular $\frac{\partial}{\partial z}v(x,z) >0$. Then, $-\int_0^\infty \frac{\partial}{\partial z}v(a_2,z) \Delta F^2(z)\dd{z} < 0$ while the rest of integrals in \eqref{eq:t1bi} are $0$, which contradicts \eqref{eq:t1bi}.

Finally, we prove the necessity of \eqref{eq:fsbd}. Assume that there exists some $x,z$ such that $F_A(x,z)-F_B(x,z)>0$. We will first show contradiction for $x<0$, but we need to define function $v$ that fulfills Definition \ref{deff:lasbdproperty} so it is defined on the whole domain of $x$. Let $x<0$. From the fact that the distribution function is right continuous, it follows that there is a neighborhood $(-b,-a)\times(c, d)$, $b>a>0, d>c>0$, such that for all $(x,z)$ in this neighbourhood, $F_A(x,z)-F_B(x,z)>0$. Consider the following function

\[
v_4(x,z)=\begin{cases}
    b(c-d)-ac & x\leq -b, z\leq c\\
    (d-c)x-ac & x\in (-b,-a), z\leq c\\
    dx & 0>x\geq -a, z\leq c\\    
    (b-a)z-bd & x\leq -b, z\in (c,d)\\
    -xz+dx-az& (x,z)\in (-b,-a)\times (c,d)\\   
    dx & 0>x\geq -a, z\in (c,d)\\
    -ad & x\leq -b, z\geq d\\
    -ad & x\in (-b,-a), z\geq d\\
    dx & 0>x\geq -a, z\geq d\\
    0 & x\geq 0.
\end{cases}
\]

Let us now check that $v_4$ fulfills Definition \ref{deff:lasbdproperty}. Firstly, for $x<0$, it is negative in each of the nine areas, which follows from $-b < - a < 0$ and $0<c<d$. It is $0$ for $x\geq 0$. Secondly, it is non-decreasing, e.g. the derivative of $(b-a)z-bd$ with respect to $z$ is $b-a>0$, or the derivative of $-xz+dx-az$ with respect to $z$ is $-x - a > 0$, because $x \in (-b, -a)$. Also, the derivative of $-xz+dx-az$ with respect to $x$ is $-z + d >0$, because then $z < d$. On the other hand, $\frac{\partial v(x, z)}{\partial x}=0$ for $x\geq 0$. Thirdly, it is loss averse, as the derivative $\frac{\partial v(-x, z)}{\partial x} > 0 = \frac{\partial v(x, z)}{\partial x}$. Finally, it is submodular given that the cross derivative of $-xz+dx-az$ is $-1$ when $(x,z)\in (-b,-a)\times (c,d)$ and $0$ elsewhere. The values were chosen so that the function is continuous at the boundaries of nine areas.

Finally, for $x>0$, the following function will lead to a contradiction in the same way as $v_4$. ``$C$ large'' below means that a constant $C$ is chosen such that its value is sufficient to ensure that the derivative of $\frac{\partial v_5}{\partial x} (-x) \geq \frac{\partial v_5}{\partial x} (x)$. 

\[
v_5(x,z)=\begin{cases}
    Cx& x\leq 0, \: C \text{ large}\\
    dx+bz-ac & 0<x\leq a, z\leq c\\
    dx-cx+bz & x\in (a,b), z\leq c\\
    dx+bz-cb & x\geq b, z\leq c\\    
    dx+bz-az & 0<x\leq a, z\in (c,d)\\
    -xz+dx+bz& (x,z)\in (a,b)\times (c,d)\\   
    dx& x\geq b, z\in (c,d)\\
    dx+bz-ad & 0<x\leq a, z\geq d\\
    bz & x\in (a,b), z\geq d\\
    dx+bz-bd & x\geq b, z\geq d\\
\end{cases}
\]

\end{proof}

\begin{proof}[Proof of Corollary \ref{corr:additive}]
    This is a direct consequence of the fact that in this case
        \[
W(F)=-\int_{-\infty}^{0} v'_1(x) F^1(x)\dd{x}+\int_{0}^{\infty} v'_1(x) (1-F^1(x))\dd{x}-\int_0^\infty v'_2(z) F^2(z)\dd{z}
\]
because $\frac{\partial^2}{\partial x\partial z}v(x,z)= 0$.
\end{proof}

\begin{proof}[Proof of Theorem \ref{thm:lasbd2}]
    Notice that $K^1(x)=K(x,0)$ and $K^2(z)=K(-a_1,z)$. Proceeding in the same way as in Theorem \ref{thm:lasbd} and substituting $\Delta K$ in \eqref{eq:t1bi} we have that
     \begin{multline}\label{eq:t1bixk}
     \Delta W =   \int_0^{\infty}\frac{\partial}{\partial x}(v(x,0)-v(-x,0)) \Delta K^1(-x)\dd{x}-\int_{0}^{\infty}\frac{\partial}{\partial x}v(x,0)(\Delta K^1(x)+\Delta K^1(-x))\dd{x}\\-\int_0^\infty \frac{\partial}{\partial z}v(-a_1,z) \Delta K^2(z)\dd{z}-\int_{-\infty}^{\infty}\int_0^\infty \frac{\partial^2}{\partial x\partial z}v(x,z)\Delta K(x,z)\dd{z}\dd{x} \geq 0. 
    \end{multline}

\noindent Given that $K^1(x)=K(x,0)=F^1(x)$ and $K^2(z)=K(-a_1,z)=F^2(z)$ and utilizing the assumptions of loss aversion, non-decreasingness given in Definition \ref{deff:lasbdproperty} and supermodularity, we have that \eqref{eq:lasd}, \eqref{eq:fsd} and \eqref{eq:fsbd2} are sufficient for \eqref{eq:t1bixk} to hold for any $v$. 

Conditions \eqref{eq:lasd} and \eqref{eq:fsd} are as in Theorem \ref{thm:lasbd}. We show the necessity of \eqref{eq:fsbd2} by contradiction. Assume that there exists some $x,z$ such that $K_A(x,z)-K_B(x,z)>0$. Let $x<0$. From the fact that the distribution function is right continuous, it follows that there is a neighborhood $(-b,-a)\times(c, d)$, $b>a>0, d>c>0$, such that for all $(x,z)$ in this neighbourhood, $K_A(x,z)-K_B(x,z)>0$. Consider the following function

\[
\tilde{v}_4(x,z)=\begin{cases}
    -bc+bz+dx-bd & x\leq -b, z\leq c\\
    (c+d)x+bz-bd & x\in (-b,-a), z\leq c\\
    -(c+d)a+bz-bd & 0>x\geq -a, z\leq c\\    
    dx-bd & x\leq -b, z\in (c,d)\\
    xz+dx+bz-bd& (x,z)\in (-b,-a)\times (c,d)\\   
    (b-a)z-(a+b)d & 0>x\geq -a, z\in (c,d)\\
    dx-bd & x\leq -b, z\geq d\\
    2xd & x\in (-b,-a), z\geq d\\
    -2ad & 0>x\geq -a, z\geq d\\
    0 & x\geq 0.
\end{cases}
\]

Let us now check that $\tilde{v}_4$ fulfills Definition \ref{deff:lasbdproperty} but with supermodularity. Firstly,  for $x<0$, it is negative in each of the nine areas, given that $-b < - a < 0$ and $0<c<d$. It is $0$ for $x\geq 0$. Secondly, it is non-decreasing, e.g. the derivative of $(c+d)x+bz-bd$ with respect to $x$ is $c+d>0$, or the derivative of $xz+dx+bz-bd$ with respect to $x$ is $z+d > 0$, because $z>0$.  Also, the derivative of $xz+dx+bz-bd$ with respect to $z$ is $x + b >0$ when $x > -b$. Thirdly, it is loss averse, as the derivative $\frac{\partial \tilde{v}_4(-x, z)}{\partial x} > 0 = \frac{\partial \tilde{v}_4(x, z)}{\partial x}$ for $x\geq 0$. Finally, it is supermodular given that the cross derivative of $xz+dx+bz-bd$ is $1$ and $0$ elsewhere. 

Now let $x>0$ and for all $x, z \in (a, b)\times (c, d)$ we have that  $K_A(x,z)-K_B(x,z)>0$. The following function with $\tilde{C}$ chosen appropriately so that loss aversion is preserved, will lead to a contradiction.

\[
\tilde{v}_5(x,z)=\begin{cases}
    \tilde{C}x& x\leq 0, \tilde{C} \text{ large}\\
    ac & 0<x\leq a, z\leq c\\
    cx & x\in (a,b), z\leq c\\
    bc & x\geq b, z\leq c\\    
    az & 0<x\leq a, z\in (c,d)\\
    xz& (x,z)\in (a,b)\times (c,d)\\   
    bz& x\geq b, z\in (c,d)\\
    ad & 0<x\leq a, z\geq d\\
    dx & x\in (a,b), z\geq d\\
    bd & x\geq b, z\geq d\\
\end{cases}
\]
\end{proof}

\begin{proof}[Proof of Theorem \ref{thm:iasd}] 
Let $v_x=\frac{\partial}{\partial x} v$; $v_z=\frac{\partial}{\partial z} v$; further, let $v_{xx}, v_{zz}$ denote respective second order derivatives and $v_{xzz}, v_{xxz}, v_{xxzz}$ mixed derivatives. To obtain second order conditions, we need to integrate \eqref{eq:svf2bi} in Lemma \ref{lem:svf2bi} by parts. Let us first concentrate on the first two terms that correspond to, respectively, losses and gains, as the integration here is less standard. Let us denote

\[
I:=-\int_{-\infty}^{0}\frac{\partial}{\partial x}v(x,a_3) F^1(x)\dd{x}+\int_{0}^{\infty}\frac{\partial}{\partial x}v(x,a_3) (1-F^1(x))\dd{x}
\]
in \eqref{eq:svf2bi}. We have that $H^{1'}(t) = F(t) \dd t$ and $S^{1'}(t)=(F(t) - 1) \dd t$. Recalling our bounded support assumption, we have

\begin{multline*}
 -\int_{-\infty}^{0}\frac{\partial}{\partial x}v(x,a_3) F^1(x)\dd{x} = -\int_{-\infty}^{0}\frac{\partial}{\partial x}v(x,a_3) \dd{H^1(x)} =  -\int_{-\infty}^{0}\frac{\partial}{\partial x}v(x,a_3) H^{1'}(x) \dd{x} =\\
 = - \frac{\partial}{\partial x}v(x,a_3) H^1(x)|_{-a_1}^0 + \int_{-\infty}^0 \frac{\partial^2}{\partial x^2}v(x,a_3) H^1(x)\dd{x} = -\frac{\partial}{\partial x}v(0,a_3) H^1(0) + \int_{-\infty}^0 \frac{\partial^2}{\partial x^2}v(x,a_3) H^1(x)\dd{x} = \\
 = -\frac{\partial}{\partial x}v(-a_1,a_3) H^1(0) + \int_{-\infty}^0 \frac{\partial^2}{\partial x^2}v(x,a_3) \rbr{H^1(x) - H^1(0)}\dd{x}
\end{multline*}

\noindent and

\begin{multline*}
\int_{0}^{\infty}\frac{\partial}{\partial x}v(x,a_3) (1-F^1(x))\dd{x} = \int_{0}^{\infty}\frac{\partial}{\partial x}v(x,a_3) (-S^{1'}(x))\dd{x} = \\
= \frac{\partial}{\partial x}v(x,a_3) (-S^1(x))|_0^{a_2} - \int_0^{\infty} \frac{\partial^2}{\partial x^2}v(x,a_3) (-S^1(x))\dd{x} =  \frac{\partial}{\partial x}v(0,a_3) S^1(0) + \int_0^{\infty} \frac{\partial^2}{\partial x^2}v(x,a_3) S^1(x)\dd{x} =\\
= \frac{\partial}{\partial x}v(a_2,a_3) S^1(0) + \int_0^{\infty} \frac{\partial^2}{\partial x^2}v(x,a_3) \rbr{S^1(x)-S^1(0)}\dd{x}
\end{multline*}

Putting these two pieces together we have

\begin{multline*}
    I = \frac{\partial}{\partial x}v(a_2,a_3)S^1(0) - \frac{\partial}{\partial x}v(-a_1,a_3)H^1(0) + \int_{-\infty}^0 \frac{\partial^2}{\partial x^2}v(x,a_3) \rbr{H^1(x) - H^1(0)}\dd{x} + \\
    + \int_0^{\infty} \frac{\partial^2}{\partial x^2}v(x,a_3) \rbr{S^1(x)-S^1(0)}\dd{x}
\end{multline*}

\noindent and

\begin{multline*}
    \Delta I = \frac{\partial}{\partial x}v(a_2,a_3)\rbr{S^1_A(0)-S^1_B(0)} - \frac{\partial}{\partial x}v(-a_1,a_3)\rbr{H^1_A(0)-H^1_B(0)} + \\
    + \int_{-\infty}^0 \frac{\partial^2}{\partial x^2}v(x,a_3) \rbr{\rbr{H^1_A(x) - H^1_A(0)} - \rbr{H^1_B(x) - H^1_B(0)}}\dd{x} + \\
    + \int_0^{\infty} \frac{\partial^2}{\partial x^2}v(x,a_3) \rbr{\rbr{S^1_A(x)-S^1_A(0)} - \rbr{S^1_B(x)-S^1_B(0)}}\dd{x},
\end{multline*}

\noindent or equivalently

\begin{multline}\label{eq:deltaI}
    \Delta I = \frac{\partial}{\partial x}v(a_2,a_3)\rbr{S^1_A(0)-S^1_B(0)} - \frac{\partial}{\partial x}v(-a_1,a_3)\rbr{H^1_A(0)-H^1_B(0)} + \\
    + \int_{-\infty}^0 \frac{\partial^2}{\partial x^2}v(x,a_3) \rbr{\rbr{H^1_A(x) - H^1_B(x)} - \rbr{H^1_A(0) - H^1_B(0)}}\dd{x} + \\
    + \int_0^{\infty} \frac{\partial^2}{\partial x^2}v(x,a_3) \rbr{\rbr{S^1_A(x)-S^1_B(x)} - \rbr{S^1_A(0)-S^1_B(0)}}\dd{x},
\end{multline}

\noindent from which we can see that given the S-shapedness of $v$ the conditions 
$H^1_A(x) - H^1_B(x) \geq H^1_A(0) - H^1_B(0)$ and $S^1_A(x) - S^1_B(x) \leq S^1_A(0) - S^1_B(0)$ for all $x \geq 0$ are sufficient for the last two terms to be positive. Furthermore, taking $x=-a_1$ in the first condition we get that $0 \geq H^1_A(0) - H^1_B(0)$ and similarly, taking $x=a_2$ in the second condition we get that $0 \leq S^1_A(0) - S^1_B(0)$, so these two conditions are sufficient for $\Delta I \geq 0$ and they are equivalent to \eqref{eq:iasd1}. Furthermore, there is also equivalence with \eqref{eq:iasd}, namely, for all $x>0>y$ we have

\begin{align*}
 S^1(x) - H^1(y) - \rbr{S^1(0)-H^1(0)} &=
 \int_x^{\infty} 1-F^1(t) \dd t - \int_{-\infty}^{y} F^1(t) \dd t - \rbr{\int_0^{\infty} 1-F^1(t) \dd t - \int_{-\infty}^{0} F^1(t) \dd t} \\    {} &= \rbr{-\int_0^x 1-F^1(t) \dd t} - \rbr{\int_{-\infty}^{y} F^1(t) \dd t - \int_{-\infty}^0 F^1(t) \dd t} \\
 {} &= -\int_0^x 1-F^1(t) \dd t + \int_{y}^0 F^1(t) \dd t = - \int_0^x \dd t + \int_{y}^x F^1(t) \dd t.
\end{align*}

\noindent Coming back to \eqref{eq:iasd1} the first term cancels out and we get \eqref{eq:iasd}.

Let us now come back to the full expression for $\Delta W$. Rewriting $\Delta I$ and further integrating \eqref{eq:svf2bi} by parts we get 
\begin{multline}\label{eq:wint2}
    \Delta W =  v_x(a_2,a_3)\Delta S^1(0)  - v_x(-a_1,a_3)\Delta H^1(0) 
    + \int_{-\infty}^0 v_{xx}(x,a_3) \rbr{\Delta H^1(x) - \Delta H^1(0)}\dd{x} + \\
    + \int_0^{\infty} v_{xx}(x,a_3) \rbr{\Delta S^1(x) - \Delta S^1(0)}\dd{x} -v_z(a_2,a_3)\Delta H^2(a_3)+
 \int_{0}^{\infty} v_{zz}(a_2,z)\Delta H^2(z)\dd{z}+v_{xz}(a_2,a_3)\Delta H(a_2,a_3)+\\
 -\int_{-\infty}^{\infty} v_{xxz}(x,a_3)\Delta H(x,a_3)\dd{x}-
 \int_{0}^{\infty} v_{xzz}(a_2,z)\Delta H(a_2,z)\dd{z}+\int_{-\infty}^{\infty}\int_{0}^{\infty} v_{xxzz}(x,z)\Delta H(x,z)\dd{z}\dd{x}\geq 0.
\end{multline}
From this we see that, apart from \eqref{eq:iasd} (or \eqref{eq:iasd1}), conditions \eqref{eq:ssd} and \eqref{eq:ssbd} are sufficient. We now show that these conditions are also necessary by means of a contradiction. In order to violate \eqref{eq:iasd} we will first assume that it is violated in the negative area. Towards a contradiction, assume that there exist intervals $(-b, -a)$, $b > a \geq 0$ such that for all $x \in (-b, -a)$ we have $H^1_A(x) - H^1_B(x) < H^1_A(0) - H^1_B(0)$, or equivalently, $\Delta H^1(x) - \Delta H^1(0) < 0$. The function $v_8$  

\[
v_8(x, z)=\begin{cases}
a^2-b^2 & x\leq-b \\
x^2+2bx+a^2 & -b < x \leq -a \\
2(b-a)x &-a < x \leq 0\\
0 & x \geq 0.
\end{cases}
\]
fulfills Definition \ref{deff:iasdproperty}. Most importantly, for all $x \in (-b, -a)$ we have $v_{xx} > 0$ and $v_{xx}=0$ otherwise. Thus using this $v_8$ we obtain $ \int_{-\infty}^{\infty} v_{xx}(x,a_3)\rbr{\Delta H^1(x) - \Delta H^1(0)}\dd{x} < 0$, and the rest of the terms in \eqref{eq:wint2} is zero, a contradiction. The case of positive area is the same, but the function $v$ should be concave instead of convex. 

In a similar fashion, towards a contradiction with \eqref{eq:ssd}, assume there exists an interval $(c,d)$, $d > c \geq 0$, such that for all $z \in (c,d)$ condition \eqref{eq:ssd} is violated: $\Delta H^2(z) > 0$. The function $v_9$, fox $x \geq 0$  

\[
v_9(x, z)=\begin{cases}
0 & z = 0 \\
2(d-c)z & 0 < z \leq c\\
-z^2 + 2dz - c^2 & c < z \leq d \\
d^2 - c^2 & d < z.
\end{cases}
\]
and $bx, b>0$ for $x<0$ fulfills Definition \ref{deff:iasdproperty}. Most importantly, for all $z \in (c, d)$ we have $v_{zz}< 0$ and $v_{zz}=0$ otherwise. Thus using $v_9$ we obtain $\int_{0}^{\infty} v_{zz}(a_2,z)\Delta H^2(z)\dd{z} <0$, while the rest of terms in \eqref{eq:wint2} is zero, a contradiction with \eqref{eq:wint2}.

The necessity of \eqref{eq:ssbd} can be shown by applying the same approach as in Theorem \ref{thm:lasbd}, but replacing $v_4$ with respective derivatives. That is, we take $v_4=\frac{\partial^2}{\partial x\partial z}v_6(x,z)$. Then, 
$\frac{\partial^4}{\partial x^2\partial z^2}v_6(x,z)=\frac{\partial^2}{\partial x\partial z}v_4(x,z)$, i.e. 
$(v_6)_{xxzz}=(v_4)_{xz}$. Similarly, we take $v_3=\frac{\partial^2}{\partial x \partial z} v_7(x, z)$ and we have that $(v_7)_{xzz}=(v_3)_{z}$. A bit more tricky is the case of $v_{xxz}$ where we take $v_1=\frac{\partial^2}{\partial x \partial z} v_8(x, z)$, and $-v_1(-x,z)=\frac{\partial^2}{\partial x \partial z} v_9(x, z)$.
We obtain $(v_8)_{xxz}=(v_1)_{x}$ on the negative domain of integration and $(v_9)_{xxz}(x,z)=-(v_1)_{x}(-x,z)$ for $x> 0$ on the positive domain of integration.
\end{proof}

\begin{proof}[Proof of Theorem \ref{thm:iasd2}]
Proceeding in the same way as in Theorem \ref{thm:iasd}, substituting $\Delta L$ into \eqref{eq:wint2} we get 

\begin{multline}\label{eq:wint2x2}
    \Delta W =v_x(a_2,0)\Delta S^1(0)  - v_x(-a_1,0)\Delta H^1(0) 
    + \int_{-\infty}^0 v_{xx}(x,0) \rbr{\Delta H^1(x) - \Delta H^1(0)}\dd{x} + \\
    + \int_0^{\infty} v_{xx}(x,0) \rbr{\Delta S^1(x) - \Delta S^1(0)}\dd{x} -v_z(-a_1,a_3)\Delta H^2(a_3)+
 \int_{0}^{\infty} v_{zz}(-a_1,z)\Delta H^2(z)\dd{z}
 -v_{xz}(a_2,a_3)\Delta L(a_2,a_3) +\\
 +\int_{-\infty}^{\infty} v_{xxz}(x,a_3)\Delta L(x,a_3)\dd{x}+
 \int_{0}^{\infty} v_{xzz}(a_2,z)\Delta L(a_2,z)\dd{z}-\int_{-\infty}^{\infty}\int_{0}^{\infty} v_{xxzz}(x,z)\Delta L(x,z)\dd{z}\dd{x} \geq 0,
\end{multline}
Conditions \eqref{eq:iasd1} and \eqref{eq:ssd} are sufficient and necessary as in Theorem \ref{thm:iasd}, and condition \eqref{eq:ssbd2} is sufficient too. 

The necessity of \eqref{eq:ssbd2} can be shown by noticing that \eqref{eq:wint2} and \eqref{eq:wint2x2} look similar, except that $\Delta H$ is replaced by $\Delta L$ and signs of the terms that include $\Delta L$ are opposite. In those terms we can consider $-v$ and see that it needs to have opposite sign, that is, if for \eqref{eq:wint2} $v$ needs to be submodular, decreasingly submodular and second-degree submodular and for \eqref{eq:wint2x2} it needs to be supermodular, decreasingly supermodular and second-degree supermodular.
\end{proof}

\begin{proof}[Proof of Theorem \ref{thm:liasd}]
Using Lemma \ref{lem:svf2bi} we have 
    \begin{multline*}
        \Delta W= \int_0^{\infty} \frac{\partial}{\partial x}(v(x,a_3)-v(-x,a_3)) \Delta F^1(-x)\dd{x}-\int_{0}^{\infty}\frac{\partial}{\partial x}v(x,a_3)(\Delta F^1(x)+\Delta F^1(-x))\dd{x}\\-\int_0^\infty \frac{\partial}{\partial z}v(a_2,z) \Delta F^2(z)\dd{z}+\int_{-\infty}^{\infty}\int_0^\infty \frac{\partial^2}{\partial x\partial z}v(x,z)\Delta F(x,z)\dd{z}\dd{x}.
    \end{multline*}
 We integrate only the last two terms by parts getting
\[
\int_0^\infty \frac{\partial}{\partial z}v(a_2,z) \Delta F^2(z)\dd{z}=\frac{\partial}{\partial z}v(a_2,a_3) \Delta H^2(a_3)-\int_0^\infty \frac{\partial^2}{\partial z^2}v(a_2,z) \Delta H^2(z)\dd{z}
\]
and
\begin{multline*}
    \int_{-\infty}^{\infty}\int_0^\infty \frac{\partial^2}{\partial x\partial z}v(x,z)\Delta F(x,t)\dd{z}\dd{x}=\\
    \frac{\partial^2}{\partial x\partial z}v(a_2,a_3)H(a_2,a_3) - \int_{-\infty}^{\infty}\frac{\partial^3}{\partial x^2\partial z}v(x,a_3)\Delta H(x,a_3)\dd{x} \\
    -\int_0^\infty \frac{\partial^3}{\partial x\partial z^2}v(a_2,z)\Delta H(a_2,z)\dd{z}\\
    +\int_{-\infty}^{\infty}\int_0^\infty \frac{\partial^4}{\partial x^2\partial z^2}v(x,z)\Delta H(x,z)\dd{z}\dd{x}
\end{multline*}
Altogether we have 
\begin{multline}\label{eq:wlsid}
  \Delta W =  \int_0^{\infty} (v_x(x,a_3)-v_x(-x,a_3)) \Delta F^1(-x)\dd{x}-\int_{0}^{\infty}v_x(x,a_3)(\Delta F^1(x)+\Delta F^1(-x))\dd{x}\\ - v_z(a_2,a_3) \Delta H^2(a_3)+\int_0^\infty v_{zz}(a_2,z) \Delta H^2(z)\dd{z} + 
    v_{xz}(a_2,a_3)H(a_2,a_3) - \int_{-\infty}^{\infty}v_{xxz}(x,a_3)\Delta H(x,a_3)\dd{x} \\
    -\int_0^\infty v_{xzz}(a_2,z)\Delta H(a_2,z)\dd{z}
    +\int_{-\infty}^{\infty}\int_0^\infty v_{xxzz}(x,z)\Delta H(x,z)\dd{z}\dd{x} \geq 0.
\end{multline}
Given the properties of function $v$ as in Definition \ref{deff:liasdproperty} we can see that the conditions \eqref{eq:lasd} (or, equivalently \eqref{eq:lasd1} and \eqref{eq:lasd2}) in Theorem \ref{thm:lasbd} and conditions \eqref{eq:ssd} and \eqref{eq:ssbd} in Theorem \ref{thm:iasd} are sufficient for \eqref{eq:wlsid} to hold. 

Necessity of condition \eqref{eq:lasd} can be shown in the same way as in Theorem \ref{thm:lasbd}. Similarly, necessity of condition \eqref{eq:ssd}  can be shown in the same way as in Theorem \ref{thm:iasd}. And finally, necessity of condition \eqref{eq:ssbd} can be shown in the same way as in Theorem \ref{thm:iasd} as well with the exception of the case of $v_{xxz}$, where we take $v_1=\frac{\partial^2}{\partial x \partial z} v_8(x, z)$, and $v_9=\begin{cases}
    \int^z\int^x -v_1(-t,s)\dd{t}\dd{s} & x>0\\
    Cx & x\leq 0
\end{cases}$, for some large $C>0$ to preserve loss aversion in $x$. We obtain $(v_8)_{xxz}=(v_1)_{x}$ on the negative domain of integration  and $(v_9)_{xxz}(x,z)=-(v_1)_{x}(-x,z)$ for $x> 0$ on the positive domain of integration.
\end{proof}

\begin{proof}[Proof of Theorem \ref{thm:liasd2}]
Proceeding in the same way as in Theorem \ref{thm:liasd} we obtain
 
\begin{multline}\label{eq:wlsidx}
  \Delta W =  \int_0^{\infty} (v_x(x,0)-v_x(-x,0)) \Delta K^1(-x)\dd{x}-\int_{0}^{\infty}v_x(x,0)(\Delta K^1(x)+\Delta K^1(-x))\dd{x}\\ - v_z(-a_1,a_3) \Delta K^2(a_3)+\int_0^\infty v_{zz}(a_2,z) \Delta H^2(z)\dd{z} - v_{xz}(a_2,a_3)L(a_2,a_3) + \int_{-\infty}^{\infty}v_{xxz}(x,a_3)\Delta L(x,a_3)\dd{x} \\
    +\int_0^\infty v_{xzz}(a_2,z)\Delta L(a_2,z)\dd{z}
    -\int_{-\infty}^{\infty}\int_0^\infty v_{xxzz}(x,z)\Delta L(x,z)\dd{z}\dd{x} \geq 0.
\end{multline}
Given the properties of function $v$ as in Definition \ref{deff:liasdproperty} with some modified as in Definition \ref{deff:liasd2} we can see that the conditions \eqref{eq:lasd} in Theorem \ref{thm:lasbd}, \eqref{eq:ssd} in Theorem \ref{thm:iasd} and \eqref{eq:ssbd2} in Theorem \ref{thm:iasd2} are sufficient and necessary for \eqref{eq:wlsidx} to hold. 
\end{proof}

\begin{proof}[Proof of Theorem~\ref{thm:consistent}]
All of the functions $g$ discussed in the main text map a pair of distribution functions to a norm of a (vector) function.  There is a common vocabulary of transformations for all the functions.  Letting $f = (f_A, f_B) \in (\ell^\infty(\R^k))^2$ (the space of pairs of bounded functions from $\R^k$ to $\R$), these common elements include the maps
\begin{equation} \label{eq:trans}
    f \mapsto f_A \pm f_B, \quad f \mapsto \int_a^b f, \quad f \mapsto \|[f]_+\|.
\end{equation}

The bootstrap depends on the Hadamard directional derivative of the map $(F_A, F_B) \mapsto \|[g(F_A, F_B)]_+\|$.  Hadamard directionally differentiable maps obey a chain rule \citep{Shapiro90} and therefore it is of interest to note the derivatives here.  The first two transforms in~\eqref{eq:trans} are linear, and thus fully (therefore also directionally) Hadamard differentiable maps. Finally, for direction $h \in \ell^\infty(\R^k)$, \citet{Firpoetal} show that when $f \leq 0$,
\begin{equation}
	\lim_{t \searrow 0} \left| t^{-1} (\|[f+h]_+\| - \|[f]_+\|) - \|[h \cdot \chi_0]_+\| \right| = 0,
\end{equation}
where $\chi_0(x) = 1$ if $f(x) = 0$ and is zero otherwise.  

Given the form of the derivatives, all the test statistics have can be composed with some chain of the above maps.  Under Assumptions~\ref{assumptionA_first} and~\ref{assumptionA_last}, we have $\sqrt{n} ((\hat{F}_A, \hat{F}_B) - (F_A, F_B)) \leadsto \mathcal{G}_F$, where $\mathcal{G}_F$ is a Gaussian process, and this convergence is uniform over the space $\mathcal{F}$ (apply Theorem 2.8.4 of \citet{vanderVaartWellner23} to the sets $I(X \leq x)$, for $x \in \R^k$).  Then Theorem 3.2 of \citet{FangSantos19} implies the result.
\end{proof}

\begin{proof}[Proof of Theorem~\ref{thm:size}]
The set of distribution functions is convex and the function $g \mapsto \|g \cdot \chi_{c_n}\|$ is convex, while the other mappings of distribution function $F$ to test statistic $T$ are linear, implying the test statistic is a convex map of the distribution functions.  Furthermore, Corollary A.2.9 of \citet{vanderVaartWellner23} implies that if $q(1 - \alpha) > 0$, the CDF of the limiting distribution of $\sqrt{n} T_n$ is strictly increasing.  Finally, it is assumed that local distributions are in $\mathcal{F}_0$.  Then Theorem~\ref{thm:consistent} and an application of Corollary 3.2 of \citet{FangSantos19} imply the result.
\end{proof}

\singlespace
\bibliographystyle{plainnat}
\bibliography{bib_LAIA}

\end{document}